\documentclass[11pt]{article}

\usepackage[bf,small]{caption}
\usepackage[export]{adjustbox}
\usepackage[sort&compress,square,numbers]{natbib}
\usepackage[title]{appendix}
\usepackage{amsmath,amsthm,amssymb,setspace,enumitem,epsfig,titlesec,verbatim,color,array,multirow,comment,graphicx}
\usepackage{amssymb,amsfonts,amsmath,amsthm}
\usepackage{mathtools}
\usepackage{authblk}
\usepackage{bbm}
\usepackage{fullpage}
\usepackage{graphbox}
\usepackage{xcolor}
\usepackage{graphicx}
\usepackage{hyperref}
\usepackage[capitalize]{cleveref}
\usepackage{lipsum}
\usepackage{multicol}
\usepackage{subcaption}
\usepackage{times}
\usepackage{nicefrac}

\usepackage[top=2.5cm,left=2.7cm,right=2.7cm,bottom=3.2cm]{geometry}

\usepackage[nomarkers,figuresonly,nofiglist]{endfloat}

\let\geq=\geqslant %

\makeatletter
\def\blfootnote{\gdef\@thefnmark{}\@footnotetext}
\makeatother

\def\FigureWatch{\textbf{Figure 1}}

\def\FigurePretty{\textbf{Figure 3}}
\def\FigureDirectedCycle{\textbf{Figure 4}}
\def\FigureTest{\textbf{Figure 4}}
\def\FigureRare{\textbf{Figure 2}}
\def\FigureIslands{\textbf{Figure 5}}

\smallskip

\titleformat{\section}{\sffamily \fontsize{12}{12}\bfseries}{\thesection}{1em}{}
\titleformat{\subsection}{\sffamily \fontsize{10}{10.5}\bfseries}{\thesubsection}{1em}{}

\newcommand{\SI}{{\bf SI}}

\title{\bfseries\sffamily \LARGE Fixation location in structured populations}
\date{}
\author[a,b,c]{David~A.~Brewster}
\author[d]{Gabor~Lippner}
\author[e]{Josef~Tkadlec}
\author[c,f]{Martin~A.~Nowak}

\affil[a]{John A. Paulson School of Engineering and Applied Sciences, Harvard University, Boston, MA~02134,~USA}
\affil[b]{Department of Molecular and Cellular Biology, Harvard University, Cambridge, MA~02138,~USA}
\affil[c]{Department of Organismic and Evolutionary Biology, Harvard University, Cambridge MA~02138,~USA}
\affil[d]{Department of Mathematics, Northeastern University, 360 Huntington Ave, Boston, MA 02115,~USA}
\affil[e]{Computer Science Institute, Charles University, Prague, Czech Republic}
\affil[f]{Department of Mathematics, Harvard University, Cambridge, MA 02138,~USA}

\begin{document}
\maketitle
\onehalfspacing

~\\[-1.1cm]

\noindent
\textbf{%
Whether a mutant has replaced an existing genotype or cultural trait can be uncovered by sampling the population.
However, an observer rarely knows the entire population structure and can rarely inspect every individual.
We introduce the notion of the fixation location in a network structured population as the location of the last wild type individual immediately before fixation.
We show the fixation location has several properties that the classical quantities in evolutionary dynamics---the fixation probability and the fixation time---lack.
Firstly, the likeliest fixation location is often the best location to sample to learn whether fixation has occurred.
Secondly, the last wild type tends to survive at low temperature locations, the locations receiving few offspring from neighbors.
Under strong selection some locations can never be the fixation location.
We prove exact identities for the fixation location, the fixation probability, and the fixation time.
The likeliest fixation location does not always tend to be at locations far from the initial mutant.
The fixation location is evidence of the origin of a mutant lineage.
Thirdly, population structures with similar fixation probabilities and fixation times can have different distributions of the fixation location.
Monitoring the last wild type at a few locations can distinguish a hypothesized structure from the true structure.
Finally, in a population of connected islands with migration, the strength of selection determines the island of the last wild type.
Our results have implications for understanding, monitoring, and mitigating extinction events in biological and social settings.
}

{\bf \blfootnote{Corresponding author e-mail address: dbrewster@g.harvard.edu}}

\section*{Significance Statement}
A new gene, species, or language often spreads across a region and displaces the original population.
The original population often survives longest in particular places.
Evolutionary biologists have long studied the success probability and the duration of a sweep but not the place where it ends.
We introduce the fixation location as the place of the last surviving individual from the original population.
The fixation location is most often a place receiving few offspring from other places.
The likeliest fixation location is often the best place to check to determine whether the new population has completely replaced the old population.
Where the sweep began and how places are connected can also be inferred from the fixation location.

\section*{Introduction}
In an evolving population, new variants spread by reproduction or by learning~\cite{fisher1937wave}.
Examples include strains of the bacterium \textit{Wolbachia} in mosquito populations~\cite{hoffmann2011successful,schmidt2017local}, mutant stem cells in intestinal crypts~\cite{lopezgarcia2010intestinal,vermeulen2013defining,nicholson2018fixation}, cultural traits~\cite{cavalli1981cultural,boyd1988culture}, and languages~\cite{abrams2003modelling}.
While a new variant spreads, the older variant often persists longest at particular places.
In New Zealand streams with introduced brown trout, a native fish persists in fragmented populations~\cite{townsend1991fragmented}.
To protect native salmonids, managers install barriers against the upstream movement of introduced trout into headwater streams~\cite{fausch2009invasion}.
Across many declining species, remnant populations persist more often at the edge of the historical range than at the center~\cite{channell2000dynamic}.
When a dominant language spreads, a minority language persists in a shrinking area~\cite{isern2014language}.

Evolution in a finite population can be described as a stochastic process of reproduction and death~\cite{fisher1923xxi,haldane1927mathematical,Wright_1931,moran1958random,Kimura1962,ewens2004mathematical}.
The Moran process is a classical model of reproduction and death in a population of fixed size~\cite{moran1958random,Moran1962,nowak2006evolutionary}.
The two central quantities of the Moran process are the fixation probability and the fixation time of a new mutant~\cite{Kimura1962,Kimura1969,Nowak2004}.
With rare mutations, the rate of evolution equals the rate of appearance of new mutants times the fixation probability~\cite{Kimura1968}.
In most natural populations, offspring replace individuals near the parent rather than anywhere in the population.

In evolutionary graph theory, the individuals of a population occupy the vertices of a graph~\cite{lieberman2005evolutionary,ohtsuki2006simple,allen2017evolutionary}.
The edges of the graph specify where the offspring of each individual can be placed.
Under neutral evolution, the probability of replacement of the individual at a vertex in one time step is proportional to the temperature of the vertex~\cite{lieberman2005evolutionary}.
By the isothermal theorem, every graph with equal temperatures at all vertices has the fixation probability of a well-mixed population~\cite{lieberman2005evolutionary}.
An advantageous mutant reaches fixation more often on an amplifier of selection and less often on a suppressor of selection than in a well-mixed population~\cite{lieberman2005evolutionary,adlam2015amplifiers,nowak2003linear,hindersin2015most}.
Graph families with arbitrarily strong amplification have been constructed~\cite{galanis2017amplifiers,pavlogiannis2018construction,goldberg2019asymptotically}.
The fixation probability also depends on the update rule~\cite{kaveh2015duality,allen2020transient,tkadlec2020limits,svoboda2024amplifiers,brewster2025mixed}.
The fixation probability~\cite{Hindersin2016,moller2019exploring,diaz2014approximating} and the fixation time~\cite{broom2010evolutionary,Hindersin2014,diaz2016absorption,hathcock2019fitness,tkadlec2019population,tkadlec2021fast,brewster2024fixation,kuo2024evolutionary,kopfova2025colonization} have been studied on many graphs and in subdivided populations~\cite{slatkin1981fixation,whitlock2003fixation,Houchmandzadeh2011,marrec2021toward,abbara2023frequent}.
Under weak selection, every graph can be analyzed in polynomial time, for frequency-dependent selection~\cite{allen2017evolutionary} and for constant selection~\cite{allen2021fixation}.
The rate of evolution~\cite{frean2013effect,allen2015molecular,tkadlec2019population}, the maintenance of diversity~\cite{brewster2025maintaining}, the mutation load~\cite{sharma2025population}, and the outcome of evolutionary games~\cite{Nowak_May_1992,ohtsuki2006simple,allen2017evolutionary} also depend on the population structure.

The fixation probability and the fixation time do not specify where the last wild type is located immediately before fixation.
We call the vertex of the last wild type, immediately before fixation, the fixation location.
On a complete graph, every vertex other than the vertex of the initial mutant is equally likely to be the fixation location.
A related quantity in probability theory is the last new vertex visited by a random walk~\cite{lovasz1993note}.
Only on cycles and on complete graphs is every vertex other than the starting vertex equally likely to be the last new vertex~\cite{lovasz1993note}.
The fixation location is different from the last new vertex, because a wild type can replace a mutant at any time before fixation.
To our knowledge, the distribution of the fixation location on graphs has not been analyzed.

An observer is a person who follows the spread of a new variant through a population.
Examples are a conservation biologist who surveys native fish in a river network and a linguist who surveys the speakers of a minority language.
An observer can rarely examine every individual of a population and must therefore choose where to sample.
For the early detection of an outbreak, monitoring the friends of randomly chosen individuals provides earlier warning than monitoring randomly chosen individuals~\cite{christakis2010social}.
The best place to sample for learning whether fixation has occurred is not known.
An observer may also want to know where a mutant lineage began.
Most methods for locating the source of a spreading process require observations made before the end of the spread~\cite{shah2011rumors,pinto2012locating,brockmann2013hidden}.
After fixation, a snapshot of the population provides no information about the initial vertex, because every vertex is occupied by a mutant.
An observer may also want to test a hypothesized population structure, even when two candidate structures have nearly equal fixation probabilities and fixation times.
The fixation location has not been used to locate the initial vertex or to distinguish population structures.

Here we study the fixation location under Birth-death updating on directed graphs.
We consider neutral evolution, constant selection of any strength, and the colonization process, where only mutants reproduce.
We ask where the last wild type is likely to be located and what an observer can learn from the fixation location.

\section*{Model}
We consider a population of $N$ individuals.
Each individual is a mutant or a wild type.
Mutants have fitness $r\geq 1$ whereas
wild types have fitness $1$.
The population structure is represented as a directed graph $G=(V, E)$ with $|V|=N$ vertices (see \FigureWatch\textbf{a}).
Each vertex of $G$ is occupied by exactly one individual.
If $G$ has an edge from vertex $x$ to vertex $y$, an offspring of the individual at $x$ can replace the individual at $y$.
We call $x$ an in-neighbor of $y$ and $y$ an out-neighbor of $x$.
An undirected graph is a directed graph with every edge bidirectional.
On a weighted graph, every edge has a positive weight.

We consider Birth-death (Bd) updating.
At each time step, an individual is chosen for reproduction with probability proportional to fitness.
The offspring replaces the individual at an out-neighbor of the parent.
On an unweighted graph, the out-neighbor is chosen uniformly at random.
On a weighted graph, the out-neighbor is chosen with probability proportional to the weight of the edge.
Initially one vertex $u \in V$ is occupied by a mutant.
We call $u$ the initial vertex or the initial mutant location.
All other vertices are initially occupied by wild types.
The initial mutant and the descendants of the initial mutant form the mutant lineage.
\textit{Neutral evolution} is the regime $r=1$.
The \textit{colonization process} is the limit $r\to\infty$, where only mutants are chosen for reproduction.

We consider graphs where the population eventually \emph{absorbs}.
Absorption is the arrival at a population of only mutants or only wild types.
After absorption, the population composition remains unchanged.
Absorption with every individual a mutant is \emph{fixation}.
Absorption with every individual a wild type is \emph{extinction}.
The fixation probability $\rho_r^G(u)$ is the probability of fixation from initial vertex $u$ on $G$, with mutant fitness $r$.
The fixation time $T_r^G(u)$ is the expected number of time steps until fixation, conditional on fixation.
A fixation trajectory is a run of the process ending in fixation.
On a fixation trajectory the last wild type is replaced at a unique time step.
Let $v\in V$ be the vertex of the last wild type immediately before the replacement.
We call $v$ the \emph{fixation location}.
The probability of fixation location $v$, conditional on fixation from initial vertex $u$, is $L_r^G(v\mid u)$.
In the figures the initial vertex is uniformly random unless stated otherwise.
For a uniformly random initial vertex, the fixation location probability of vertex $v$ is $\sum_{u}\rho_r^G(u)L_r^G(v\mid u)/\sum_{u}\rho_r^G(u)$.
The likeliest fixation location is the vertex with the largest fixation location probability.

An observer of a population sees the individuals at some of the vertices.
An observer monitors a vertex by seeing the type of the individual at the vertex at every time step.
An observer samples a vertex by seeing the type of the individual at the vertex at one time step.
The sampling time is the number of time steps from the initial mutant to the sample.
Let $f_t$ be the probability of fixation by time step $t$, conditional on fixation.
An observer can see many independent fixation trajectories over time.
We call each observed fixation trajectory a fixation event.

Under neutral evolution, every individual is chosen for reproduction with probability $1/N$ at each time step.
Let $d^+(y)$ be the number of out-neighbors of vertex $y$.
An offspring of the individual at $y$ replaces the individual at each out-neighbor of $y$ with probability $1/d^+(y)$.
The temperature of vertex $v$ is $\theta_v \coloneqq \sum_{y\to v}1/d^+(y)$~\cite{lieberman2005evolutionary}.
The sum is over the in-neighbors $y$ of $v$.
On a weighted graph, the term $1/d^+(y)$ in the definition of the temperature is instead the weight of the edge from $y$ to $v$ divided by the total weight of the edges out of $y$.
Under neutral evolution, the individual at $v$ is replaced with probability $\theta_v/N$ at each time step.
The mean time between two replacements of the individual at $v$ is therefore $N/\theta_v$ time steps.
If every vertex has an out-neighbor, the average temperature over all vertices is $1$.
A low (high) temperature vertex receives offspring less (more) often than the average vertex.

\section*{Results}

\subsection*{The likeliest fixation location is often the place to sample}
Oftentimes, someone trying to observe an evolving population cannot see the whole population due to resource constraints.
So the observer must choose where to look.
Suppose the observer wants to know whether fixation has occurred.
If fixation has occurred, every vertex is occupied by a mutant.
A wild type at a sampled vertex therefore rules out fixation with certainty.
A mutant at a sampled vertex is possible with or without fixation.
The observer learns the most at the vertex where a wild type is most likely to remain.

Consider an observer who samples one vertex $w$ at sampling time $t$.
Let $m_w(t)$ be the probability of a mutant at $w$ at time step $t$, conditional on fixation.
Fixation by time step $t$ requires a mutant at $w$.
By Bayes' rule, given a mutant at $w$, the probability of fixation by time step $t$ increases from $f_t$ to
\begin{equation}\label{eq:sample}
\frac{f_t}{m_w(t)}.
\end{equation}
Given no fixation by time step $t$, the individual at $w$ is a wild type with probability $(1-m_w(t))/(1-f_t)$.
At a fixed sampling time, the two probabilities depend on the vertex only through $m_w(t)$.
The best vertex to sample at time step $t$ is therefore the vertex with the smallest $m_w(t)$.
The probability $m_w(t)$ depends on the sampling time, but the observer often does not know the sampling time.
Fortunately, the fixation location probabilities do not depend on the sampling time.
The likeliest fixation location is the vertex where the last wild type most likely remains.
We therefore compare the likeliest fixation location with the best vertex to sample.

\FigureWatch\textbf{a} displays a population structure with two regions, shaded purple and green.
Most directed edges between the two regions point from the purple region to the green region.
An observer samples four vertices at several times during one fixation trajectory at $r=1$.
A mutant occupies the pentagon vertex at the second displayed time, long before fixation.
The last wild type occupies the pentagon vertex from a later time until fixation.
\FigureWatch\textbf{b} displays the fixation location probability of every vertex at $r=1$ and at $r=5$.
The likeliest fixation location is the pentagon vertex in the green region at $r=1$.
The likeliest fixation location is the triangle vertex in the purple region at $r=5$.
\FigureWatch\textbf{c,d} displays the probability of a wild type given no fixation at each sampled vertex, against the sampling time.
In \FigureWatch\textbf{c}, the probability of a wild type given no fixation is largest at the pentagon vertex at almost every sampling time.
In \FigureWatch\textbf{d}, the probability of a wild type given no fixation is largest at the triangle vertex at almost every sampling time.
The likeliest fixation location is therefore nearly always the place to sample at $r=1$ and at $r=5$.
The place to sample depends on the mutant fitness.
On most small graphs, the likeliest fixation location is also the best vertex for a single observation (see \SI{} for details).

\subsection*{The last wild type tends to survive on low temperature vertices}
Under neutral evolution, the temperature of a vertex determines how often the vertex receives offspring.
A wild type at a low temperature vertex is rarely replaced.
We therefore expect the last wild type to occupy a low temperature vertex.
In each population structure of \FigureRare\textbf{a--d}, the fixation location probability is high where the temperature is low.
On stars, the hub has temperature $N-1$ and each leaf has temperature $1/(N-1)$ (\FigureRare\textbf{a}).
A leaf receives offspring only from the hub.
A wild type at the hub therefore always coexists with a wild type at a leaf.
The last wild type always occupies a leaf.
On paths, the mutants always occupy one block of consecutive vertices (\FigureRare\textbf{b}).
The last wild type therefore always occupies one of the two end vertices.
The two end vertices have the lowest temperature on the path.
\FigureRare\textbf{e} displays every vertex of every connected undirected graph with $N=7$ vertices.
On average, the fixation location probability of a vertex decreases with the temperature of the vertex.
The trend holds on nearly every small undirected graph with unequal temperatures (see \SI{} for details).
On the river network of the Charles River basin~\cite{lehner2013global}, the last wild type almost always occupies a low temperature vertex at a headwater (see \SI{} for details).
Directed edges can break the trend under neutral evolution (see \SI{} for details).

An exact identity holds between the fixation location, the temperatures, and the fixation probability.
The mutant set is the set of vertices currently occupied by mutants.
An active step is a time step where the mutant set changes.
Let $n_u(X)$ be the expected number of active steps from the mutant set $X$, for a single initial mutant at $u$.
We call $n_u(X)$ the escape count of $X$.
Suppose $G$ has no self-loops, every vertex has an out-neighbor, and the population eventually reaches fixation or extinction from every mutant set.
In the mutant set $V\setminus\{v\}$ the only wild type occupies $v$.
Mutant offspring arrive at $v$ at a rate proportional to $r\theta_v$.
The wild type at $v$ reproduces at a rate proportional to $1$ and always replaces a mutant.
An active step from $V\setminus\{v\}$ is therefore fixation at $v$ with probability $r\theta_v/(r\theta_v+1)$.
For every such graph, every $r\geq1$, and all vertices $u$ and $v$,
\begin{equation}\label{eq:factorization}
\rho_r^G(u)\,L_r^G(v\mid u)=n_u\bigl(V\setminus\{v\}\bigr)\,\frac{r\theta_v}{r\theta_v+1}
\end{equation}
(see \SI{} for details).
Summing Eq.~\ref*{eq:factorization} over all vertices $v$ yields the fixation probability $\rho_r^G(u)$.
The $N$ escape counts $n_u(V\setminus\{v\})$ therefore determine the fixation probability and the fixation location.

The final phase of fixation is the time with exactly one wild type.
If, in addition, every vertex has an in-neighbor and $\rho_r^G(u)>0$, then conditional on fixation from $u$ the expected number of time steps in the final phase is
\begin{equation}\label{eq:endgame}
T^{\mathrm{end}}(u)=\frac{r(N-1)+1}{r}\sum_{v\in V}L_r^G(v\mid u)\,\frac{\rho(V\setminus\{v\})}{\theta_v}
\end{equation}
(see \SI{} for details).
The probability $\rho(V\setminus\{v\})$ is the fixation probability from the mutant set $V\setminus\{v\}$.
The final phase is therefore long when the fixation location has a low temperature and $\rho(V\setminus\{v\})$ is close to $1$.
From the mutant set $V\setminus\{v\}$, fixation has probability at least $r\theta_v/(r\theta_v+1)$.
The fixation time $T_r^G(u)$ is therefore at least $(r(N-1)+1)\sum_vL_r^G(v\mid u)/(r\theta_v+1)$.
On nearly every small undirected amplifier of selection, the last wild type tends to occupy a low temperature vertex (see \SI{} for details).
The resulting long final phase is one source of the long fixation times of amplifiers of selection~\cite{tkadlec2019population}.

Some vertices are never the fixation location (\FigurePretty).
A vertex $v$ is a possible fixation location if $L_r^G(v\mid u)>0$.
By Eq.~\ref*{eq:factorization}, a vertex $v$ with an in-neighbor is a possible fixation location if and only if $n_u(V\setminus\{v\})>0$.
The escape count $n_u(V\setminus\{v\})$ is positive if and only if the mutant set $V\setminus\{v\}$ occurs with positive probability.
Consider the colonization process where a mutant is never replaced by a wild type.
The initial mutant location therefore cannot be the fixation location (\FigurePretty).
On an undirected graph, a cut vertex---a vertex whose removal disconnects the graph---cannot be the fixation location either.
The mutant lineage reaches the cut vertex before the part of the graph beyond the cut vertex.
Every other vertex can be the fixation location (\FigurePretty).
Consider the colonization process on a directed graph.
Let $G\setminus\{t\}$ be the graph $G$ without the vertex $t$ and without the edges of $t$.
A vertex $t\neq u$ is a possible fixation location exactly when $t$ has an in-neighbor and every other vertex is reachable from $u$ in $G\setminus\{t\}$.
The possible fixation locations of the colonization process are therefore computable in polynomial time on every directed graph (see \SI{} for details).

\subsection*{The fixation location is not always the farthest vertex from the initial mutant}
A natural guess for the fixation location is the vertex farthest from the initial mutant.
We show that this guess does not hold on one of the simplest graphs, the directed cycle (\FigureDirectedCycle\textbf{a}).
Every vertex has one directed edge to the next vertex around the cycle.
Every vertex therefore has temperature $1$.
Thus the temperature cannot be used to determine the fixation location on the directed cycle.
The initial mutant occupies vertex $1$.
The farthest vertex from the initial mutant is vertex $N$.
The mutants form one block of consecutive vertices.
The block grows at the front and shrinks at the back.
At each active step the front advances with probability $r/(1+r)$.
Otherwise the back retreats.
The fixation location is the vertex where the front of the block meets the back.
Let $H$ be the number of advances of the front until fixation, conditional on fixation.
The fixation location is vertex $1+H$, with vertex labels taken modulo $N$.
The distribution of $H$ has an exact formula for every $N$ and every $r$ (see \SI{} for details).

Under neutral evolution the fixation location probability of every vertex is $1/N+O(N^{-1/2}e^{-\sqrt{N/2}})$.
The fixation location is therefore nearly uniform on large cycles (\FigureDirectedCycle\textbf{b}).
Under selection the distribution of $H$ is approximately normal on large cycles (see \SI{} for details).
The mean of $H$ differs from
\begin{equation}\label{eq:cycle}
m_N\coloneqq(N-1)\frac{r}{r-1}-\frac{r+1}{(r-1)^{2}}
\end{equation}
by a term exponentially small in $N$.
The variance of $H$ is $\sigma_r^2N+O(1)$, with $\sigma_r^2\coloneqq r(r+1)/(r-1)^3$.
The fixation location therefore concentrates within order $\sqrt{N}$ vertices of vertex $1+m_N$, modulo $N$ (\FigureDirectedCycle\textbf{c,d}).
The first term of $m_N$ is the number of advances if the front advances exactly $r$ times as often as the back retreats.
The second term of $m_N$ is the effect of the conditioning on fixation.
For $r=3$, vertex $1+m_N$ is near vertex $N/2$, opposite the initial mutant (\FigureDirectedCycle\textbf{d}).
As $r\to\infty$, vertex $1+m_N$ tends to the farthest vertex $N$.
Only in the colonization process is the farthest vertex always the fixation location (see \SI{} for details).

The inverse problem to the fixation location is to find the initial mutant location from knowledge about the fixation location.
Suppose the initial mutant arises at a uniformly random vertex.
By Bayes' rule, given the fixation location $v$, the initial vertex is $u$ with probability
\begin{equation}\label{eq:origin}
\frac{\rho_r^G(u)\,L_r^G(v\mid u)}{\sum_{w\in V}\rho_r^G(w)\,L_r^G(v\mid w)}.
\end{equation}
The best guess of the initial vertex is the vertex $u$ with the largest $\rho_r^G(u)\,L_r^G(v\mid u)$.
On the directed cycle in the colonization process, the fixation location determines the initial vertex.
The vertex farthest from the fixation location is not always the best guess (see \SI{} for details).

\subsection*{The fixation location distinguishes population structures}
Two population structures can have nearly equal fixation probabilities and fixation times but can still differ in the fixation location (\FigureTest\textbf{e}).
Suppose an observer can monitor a real population but does not know its true structure.
The observer monitors a subset of the vertices (due to resource constraints) over many fixation events of the true structure.
After each fixation event, the observer notes the monitored vertex last occupied by a wild type---the observation of the fixation event.
A mutant at every monitored vertex does not rule out wild types at the other vertices.
Let $T$ be the expected fixation time for a uniformly random initial vertex, conditional on fixation.
By Markov's inequality, the probability of no fixation by time step $t$ is at most $T/t$, conditional on fixation.
The observer therefore waits a large multiple of the larger value of $T$ to have more confidence that fixation has occurred (see \SI{} for details).
The observer can then calculate the probability of each observation, empirically for the real structure and computationally for the theorized structure (\FigureTest\textbf{f}).
The observed fractions are close to the predictions of the true structure.
Let $P(x)$ and $Q(x)$ be the probabilities of observation $x$ under the true and theorized structures, respectively.
After $n$ fixation events with observations $x_1,\ldots,x_n$, the probability of the observations under the true structure is $\prod_{i=1}^nP(x_i)$.
The observer chooses the structure with the larger probability of the observations.
The probability of a wrong choice is at most $B^n$, with $B\coloneqq\sum_x\sqrt{P(x)Q(x)}$ (see \SI{} for details).
The number $B$ is below $1$ whenever $P\neq Q$.
The probability of a wrong choice therefore decreases exponentially with the number of fixation events.
With two monitored vertices, the observer chooses the true structure with high probability after a handful of fixation events (\FigureTest\textbf{g}).
With every vertex monitored, the observer needs even fewer fixation events.
An observer of only fixation times or only of the outcomes fixation and extinction needs orders of magnitude more events (\FigureTest\textbf{g}).
The fixation location can therefore distinguish population structures with nearly equal fixation probabilities and fixation times.

\subsection*{The strength of selection determines the island of the last wild type}
\FigureIslands\textbf{a} displays a population on two islands with migration between the islands.
Each island is a well-mixed population of $n$ individuals.
The offspring of an individual on island $1$ migrates to island $2$ with probability $m_{12}$.
The offspring of an individual on island $2$ migrates to island $1$ with probability $m_{21}$.
The initial mutant arises on island $1$.
Let $p_1$ be the probability of a fixation location on island $1$.

The migration probabilities can determine the island of the last wild type (\FigureIslands\textbf{c}).
Let $x_1$ and $x_2$ be the fractions of mutants on island $1$ and on island $2$.
Under neutral evolution the fixation probability is $\rho(x_1,x_2)=(m_{12}x_1+m_{21}x_2)/(m_{12}+m_{21})$ (see \SI{} for details).
The fixation probability of a single mutant on island $2$ is therefore $m_{21}/m_{12}$ times the fixation probability of a single mutant on island $1$.
In the colonization process the wild types on island $1$ are replaced at a rate proportional to $1-m_{12}+m_{21}$ near fixation.
The wild types on island $2$ are replaced at a rate proportional to $1-m_{21}+m_{12}$.
As the island size $n$ increases, $p_1$ in the colonization process tends to $1$ when $m_{12}>m_{21}$ and to $0$ when $m_{12}<m_{21}$ (see \SI{} for details).
Under neutral evolution, the exact computation of \FigureIslands\textbf{c} gives island $1$ approximately when $|m_{12}-1/2|>|m_{21}-1/2|$.
For the migration probabilities of \FigureIslands\textbf{b}, the neutral rule gives island $1$.
The colonization rule gives island $2$.
The two rules are consistent with the switch between the islands in \FigureIslands\textbf{b}.

The unbiased fitness $r^*$ is the mutant fitness with $p_1=1/2$.
The unbiased fitness depends on the population size (\FigureIslands\textbf{d}).

An observer samples one individual from one island in hopes of determining whether fixation has occurred (\FigureIslands\textbf{e}).
Let $m_i(t)$ be the mean fraction of mutants on island $i$ at time step $t$, conditional on fixation.
As in Eq.~\ref*{eq:sample}, given a mutant in the sample from island $i$, the probability of fixation by the sampling time increases from $f_t$ to $f_t/m_i(t)$.
The better island to sample is therefore the island with the smaller $m_i(t)$.
The better island to sample is island $1$ below a threshold fitness $r_s$ and island $2$ above $r_s$.
The threshold fitness $r_s$ is below the unbiased fitness $r^*$.
For mutant fitness between $r_s$ and $r^*$, the likelier island of the last wild type is the worse island to sample.
The mutant lineage arrives on island $2$ late, because offspring rarely migrate from island $1$ to island $2$.
For most of the process, island $2$ therefore has the smaller fraction of mutants.
The fixation location depends only on the end of the process.
The threshold fitness increases with the sampling time (\FigureIslands\textbf{f}).
At every sampling time of \FigureIslands\textbf{f}, the threshold fitness is below the unbiased fitness.

\section*{Discussion}
We studied where the last wild type is located immediately before fixation in a structured population.
For an observer who wants to learn whether fixation has occurred, the likeliest fixation location is the best place to sample on most small graphs.
On nearly every small undirected graph, the last wild type tends to occupy a vertex of low temperature, where a wild type is rarely replaced.
For a broad class of graphs, exact identities hold between the fixation location probabilities, the fixation probability, and the duration of the final phase.
On the directed cycle, the position of the likeliest fixation location depends on the strength of selection.
The fixation location provides information about the initial vertex and distinguishes population structures with nearly equal fixation probabilities and fixation times.
In a population of connected islands, the better island to sample is not always the likelier island of the last wild type.
An estimate of the fixation probability requires the outcomes of many trajectories, including trajectories ending in extinction.
A measurement of the fixation time requires the time of origin of each mutant.
An observation of the fixation location requires no trajectories ending in extinction and no time of origin.

Our observer samples a vertex at a single time or records only the last monitored vertex occupied by a wild type.
An observer who monitors a vertex over time sees a time series of types at the vertex.
The time series contains more information than a single sample, for example the time since the individual at the vertex became a mutant.
The best vertex to monitor with a full time series is an open question.
The choice of components to monitor is also central to the classification of perturbations in complex networks, a different setting than our paper.
In complex networks, the most sensitive or high-activity locations are not always the best components to monitor~\cite{brewster2026classifying}.

Another direction is the more general inverse problem.
Suppose an observer knows the population structure and the mutant set at some time before fixation.
The observer wants to infer the vertex of the initial mutant from the mutant set.
In many epidemiological models, an infected individual never becomes susceptible again~\cite{kermack1927contribution}.
In the Moran process, by contrast, a mutant can be replaced by a wild type.
The fixation location is the terminal case of the inverse problem, where the observed mutant set is $V\setminus\{v\}$ for the fixation location $v$.

Several natural generalizations remain open.
We assumed Birth-death updating and a constant fitness of each individual.
Under death-Birth updating, every individual dies at an equal rate~\cite{kaveh2015duality}.
The temperature of a vertex therefore yields no prediction for the fixation location under death-Birth updating.
The fixation location under frequency-dependent selection~\cite{ohtsuki2006simple,allen2017evolutionary} is also unknown.
When fitness depends on the vertex~\cite{maciejewski2014environmental,kaveh2019environmental}, vertices where the wild type has higher fitness may be natural candidates for the fixation location.
The vertex of the last mutant immediately before extinction is the counterpart of the fixation location for a mutant lineage ending in extinction.

\section*{Materials and Methods}
On small graphs, we computed the fixation probability, the fixation time, and the fixation location probabilities exactly from the Markov chain on mutant sets.
On the directed cycle and on two islands, we computed exact results from Markov chains on the mutant block and on the mutant counts.
On the river network of the Charles River basin, we estimated the fixation location probabilities from simulated fixation events.
We checked the formal statements of the \SI{} with the Lean proof assistant and the Mathlib library~\cite{demoura2021lean4,mathlib2020}.
The computational methods and mathematical proofs for our results are in the \SI{}.

{\small

\hfill\\
\noindent\textbf{AI statement.}
We used the artificial intelligence (AI) tools ChatGPT (OpenAI models GPT-6 Sol, GPT-6 Astra, and GPT-6.2 Astra), Claude (Anthropic models Claude Fable 5.2 and Claude Opus 5.5), and Antigravity (Google model Gemini 4 Neon).
The AI tools helped with writing and editing the text, with the code for the computations and the figures, with the Lean proofs, and with checking the mathematical derivations.
We did substantial work on the exposition and the design, chose the directions of the research, and checked all results.
We reviewed all output of the AI tools and take full responsibility for the results of the paper. \\

\noindent\textbf{Code availability.}
All simulations were performed in \texttt{Python} and \texttt{C++}.
All numerical calculations were performed using \texttt{Python}.
All figures were produced in part using \texttt{Python}.
Our code is available at the following web address: \url{https://github.com/davidb2/fixation-location/}. \\

\noindent
{\bf Acknowledgments.} \\
We thank David Haig and anonymous reviewers for helpful comments.
D.A.B. was supported by a Harvard Graduate School of Arts and Sciences Prize Fellowship.
J.T. was supported by GAČR grant 25-17377S and by Charles Univ. projects UNCE 24/SCI/008 and PRIMUS 24/SCI/012.
\\

\noindent
{\bf Author contributions.}\\
All authors conceived the study, performed the analysis, discussed the results, and wrote the manuscript. \\

\noindent
{\bf Competing interests.}\\
The authors declare no competing interests.\\

{\setlength{\bibsep}{0\baselineskip}
\bibliographystyle{unsrt}
\bibliography{sources}

}
}

\end{document}


\renewcommand{\thefigure}{S\arabic{figure}}
\newcommand{\figtoc}[1]{\addcontentsline{toc}{subsection}{Figure~\thefigure.\ #1}}
\renewcommand{\thetable}{S\arabic{table}}

\maketitle
\tableofcontents
\clearpage

\section{Which location to sample to learn whether fixation has occurred}\label{sec:watching}

Suppose there is a population where a replacement event is in progress.
A mutant is spreading in an initial population of all wild types.
The mutant has fitness $r\geq1$, whereas the wild type has fitness $1$.
We consider Birth-death (Bd) updating.
The population structure is represented by a (possibly directed) graph where at each time step exactly one individual resides at each vertex (location).
Our goal is to determine whether or not fixation has occurred yet.
Of course, looking at the entire population at every step, we can inspect each of the vertices and only declare fixation if all of the vertices have mutants.
But in reality, we may not have access or the resources to view all of the vertices of a graph at each step.
Thus, we consider the most restricted case when we can only sample one fixed vertex at an unknown sampled time.
In this case, the best location to sample is often the likeliest fixation location.
Next, we explain what ``best'' means and why the likeliest fixation location is often the best site to sample.

We are given access to the inhabitant of a vertex at some unknown sampled time $t$.
If we see a wild type individual at this vertex, we exclaim \texttt{fixation has NOT occurred}.
If we see a mutant individual at this vertex, we exclaim \texttt{fixation has occurred}.
What is the probability that we exclaimed correctly?
If we see a wild type individual, then it is impossible that fixation has occurred.
So that exclamation is always correct.
However, it is possible to see a mutant individual but fixation has not occurred.
Let $f_t$ be the probability of fixation by time step $t$, conditional on eventual fixation.
Let $m_w(t)$ be the probability of a mutant at vertex $w$ at time step $t$, conditional on eventual fixation.
Given that fixation has not occurred, the individual at vertex $w$ is a mutant with probability $1-(1-m_w(t))/(1-f_t)$.
Since at a fixed time step the factor $1-f_t$ is the same for every vertex,
the vertex $w$ that minimizes $m_w(t)$ minimizes the exclamation error probability.
Thus the ``best'' location to sample is the location that minimizes $m_w(t)$.
Similarly, given that a mutant is observed at vertex $w$, the probability that fixation has occurred is $f_t/m_w(t)$ by Bayes' rule.
Since at a fixed time step the factor $f_t$ is the same for every vertex,
the vertex $w$ that minimizes $m_w(t)$ maximizes the probability that the exclamation is correct.

Notice that $m_w(t)$ depends on the time elapsed.
However, we do not know the time elapsed.
We empirically show that the fixation location---a statistic independent of time---serves well as the best vertex to sample.

Consider the $850$ connected undirected graphs with $N=7$ vertices, without the $3$ graphs with all vertices equivalent.
At $r=2$ and at the expected fixation time $T$, the likeliest fixation location is an optimal vertex for a single observation on $91\%$ of the $850$ graphs.
At the same time step, the likeliest or the second likeliest fixation location is an optimal vertex for a single observation on $96\%$ of the $850$ graphs.
Even when observing the population at various times (at the expected time or the first times when $f_t$ is $0.25$, $0.5$, $0.75$, or $0.9$) and among various mutant fitness values ($r=1.25,2,\text{ and } 10$), the likeliest fixation location is an optimal vertex for a single observation on a large share of the graphs.

On larger graphs the likeliest fixation location is also usually an optimal vertex for a single observation.
Consider the $261{,}073$ connected undirected graphs with $N=9$ vertices, without the $7$ graphs with all vertices equivalent.
At the expected fixation time, the likeliest fixation location is an optimal vertex for a single observation on $90\%$, $89\%$, and $93\%$ of the $261{,}073$ graphs at $r=1.25$, $2$, and $10$, respectively.

\section{Fixation location, fixation probability, and fixation time}\label{sec:factorization}

We describe the Birth-death process by a Markov chain on the sets of mutants.
We call the set of vertices occupied by mutants the mutant set.
A time step is \emph{active} if the step changes the mutant set.
An active step from a mutant set $X\subseteq V$ is an active step with mutant set $X$ before the step.
Suppose the process starts with a single mutant at vertex $u$.
Let $n_u(X)$ be the expected number of active steps from $X$.
We call $n_u(X)$ the escape count of $X$.
Let $d^+(y)$ be the number of outgoing edges of $y$ and $\theta_v=\sum_{y\to v}1/d^+(y)$ the temperature of $v$.
Let $\rho(X)$ be the fixation probability from mutant set $X$, with $\rho(\{u\})=\rho_r^G(u)$.

\begin{proposition}[Factorization of the fixation location, Lean \Cref{lean:factorization}]\label{prop:factorization}
Let $G$ be a directed graph without self-loops on which the process is absorbed with probability one from every mutant set, for example a strongly connected graph.
For every $r\geq1$ and all vertices $u$ and $v$,
\begin{equation}
\rho_r^G(u)\,L_r^G(v\mid u)=n_u\bigl(V\setminus\{v\}\bigr)\,p_v,
\qquad
p_v=\frac{r\theta_v}{r\theta_v+\mathbbm 1[d^+(v)\geq 1]},
\end{equation}
with $p_v\coloneqq0$ when $\theta_v=0$.
\end{proposition}

\begin{proof}
From the mutant set $V\setminus\{v\}$ the active steps are of two kinds.
Either a mutant in-neighbor $y$ of $v$ places an offspring on $v$ with total weight $r\theta_v$, or
the wild type at $v$ places an offspring on an out-neighbor of $v$.
Every out-neighbor of $v$ is occupied by a mutant.
The total weight of the steps of the second kind is therefore $1$ if $v$ has an out-neighbor and $0$ otherwise.
So each active step from $V\setminus\{v\}$ ends in fixation at $v$ with probability $p_v$, independently of the past.
Fixation at $v$ happens exactly when an active step from $V\setminus\{v\}$ is of the first kind.
At most one step of the first kind occurs.
Let $S$ be the number of times the mutant set changes from $V\setminus\{v\}$ to $V$.
The number $S$ is therefore $0$ or $1$.
The event $S=1$ is the event of fixation at $v$.
Hence $\mathbb{E}[S]=P(S=1)=\rho_r^G(u)\,L_r^G(v\mid u)$.
Number the active steps from $V\setminus\{v\}$ by $k=1,2,\dots$ in order of occurrence.
Whether the $k$th active step from $V\setminus\{v\}$ occurs depends only on the process before the step.
Given the occurrence of the $k$th active step from $V\setminus\{v\}$, the step is of the first kind with probability $p_v$.
Summing over $k$ yields
\begin{align}
\mathbb{E}[S]&=
\mathbb{E}\sum_{k\geq1}\mathbbm1\big[\text{the $k$th active step from $V\setminus\{v\}$ results in $V$}\big] \\
&=
p_v\sum_{k\geq1}P\bigl(\text{the $k$th active step from $V\setminus\{v\}$ occurs}\bigr) \\
&=p_v \cdot n_u\bigl(V\setminus\{v\}\bigr)
.
\end{align}
Equating the two expressions for $\mathbb{E}[S]$ yields the identity.
\end{proof}

Summing over $v$ yields $\rho_r^G(u)=\sum_v n_u(V\setminus\{v\})\,p_v$.
If every vertex has an out-neighbor, the expected number of active steps spent one step before fixation is
\begin{equation}
\sum_{v} n_u\bigl(V\setminus\{v\}\bigr)=\rho_r^G(u)\Bigl(1+\frac1r\sum_v \frac{L_r^G(v\mid u)}{\theta_v}\Bigr).
\end{equation}
We tested the low temperature rule of the main text on all small graphs.
The relative inverse temperature of the fixation location on a graph is a ratio.
The numerator is the mean of $1/\theta_v$ over the fixation location, with the initial vertex uniformly random and conditional on fixation.
The denominator is the mean of $1/\theta_v$ over all vertices.
Consider the $273{,}183$ connected undirected graphs with $N=5$ to $9$ vertices at $r=1.25$, $2$, and $10$.
The relative inverse temperature exceeds $1$ on every non-regular graph except four graphs.
The four graphs have fixation probabilities within $0.6\%$ of the well-mixed value.
Under neutral evolution the relative inverse temperature is below $1$ on $325$ of the $273{,}183$ graphs, by less than $3\%$.
Consider the $5{,}031$ strongly connected directed graphs with $N=5$ vertices and unequal temperatures.
The relative inverse temperature is below $1$ on $53\%$ of the graphs at $r=1$ and on $2\%$ at $r=10$.
Figure~2\textbf{e} of the main text displays the fixation location probability against the temperature at $r=2$ for every vertex of every connected undirected graph with $N=7$ vertices.

\begin{figure}
  \centering
  \includegraphics[width=0.95\textwidth]{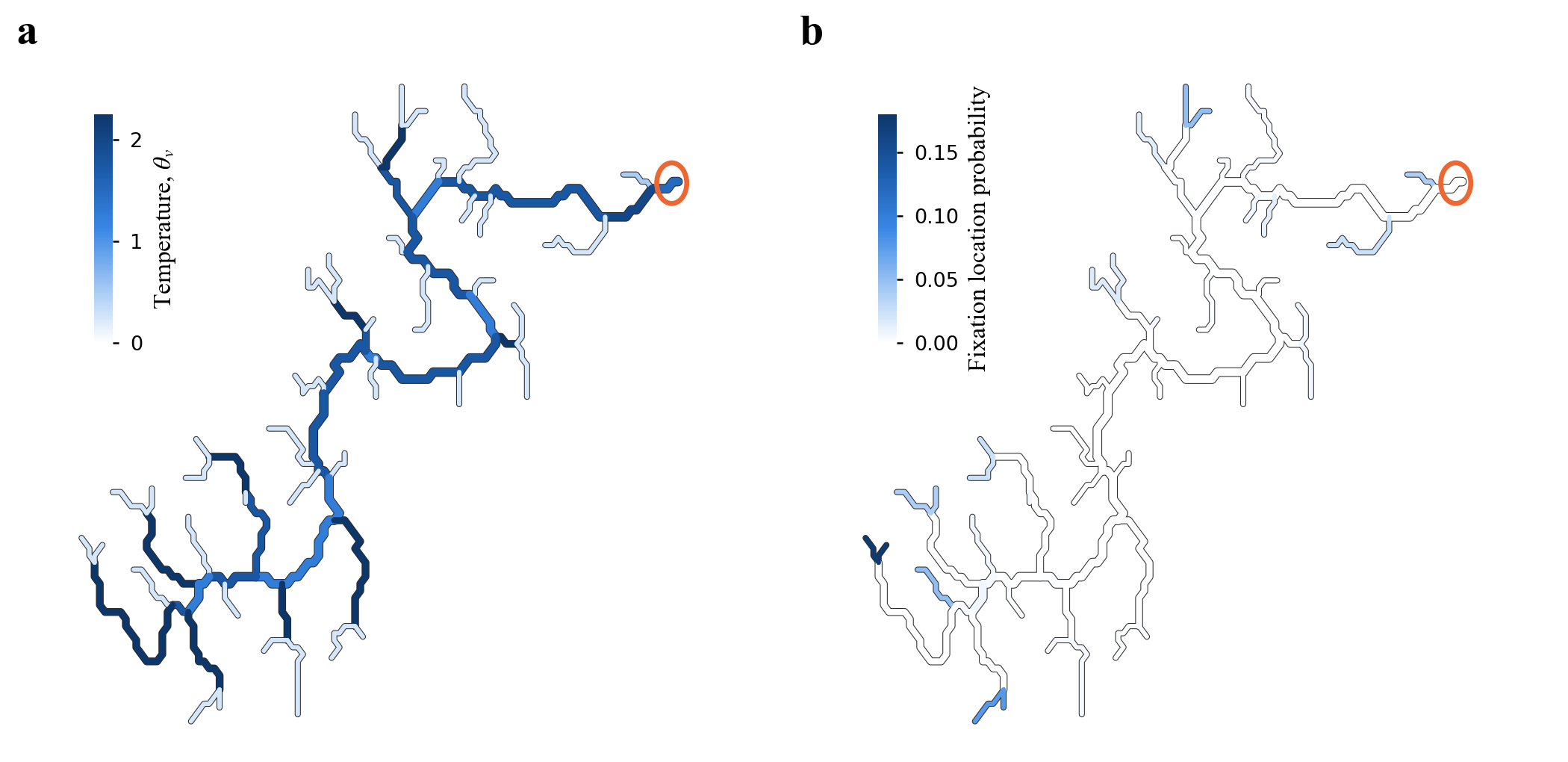}
  \caption{
    The Charles River basin as a population structure.
    Each line is one reach of the river from the HydroRIVERS database.
    Each reach is a vertex.
    Offspring are placed downstream with weight $2$ and upstream with weight $1$.
    The orange circle marks the vertex at the mouth of the river.
    \textbf{a,} Temperature $\theta_v$ of each vertex.
    \textbf{b,} Fixation location probability at $r=2$ for a uniformly random initial vertex, conditional on fixation, estimated from $4{,}000$ simulated fixation events from each initial vertex.
  }
  \label{fig:charles}
  \figtoc{The Charles River basin as a population structure}
\end{figure}

The low temperature rule also holds on an empirical population structure.
We built the river network of the Charles River basin in Massachusetts from the HydroRIVERS database~\cite{lehner2013global}.
HydroRIVERS divides every river into reaches, the stretches of river between two confluences.
Each of the reaches of the basin is a vertex ($N=77$).
The graph is a tree rooted at the vertex at the mouth of the river.
The leaves of the tree are the headwater reaches.
Water flows from the leaves toward the root.
Two vertices are joined by an edge in each direction if the two reaches meet.
The two directions have different weights.
An offspring is placed downstream, toward the root, with weight $2$ and upstream, away from the root, with weight $1$.
We give the downstream direction the larger weight because the current carries offspring downstream.
Every leaf has temperature $1/4$, except the leaf adjacent to the root with temperature $1/2$.
Every other vertex has temperature at least $5/4$.
The graph has too many vertices for the exact computation of \Cref{sec:computation}.
We therefore estimated the fixation location probabilities from $4{,}000$ simulated fixation events from each initial vertex at $r=1.5$, $2$, and $5$.
Each initial vertex has a weight proportional to the estimated fixation probability, the fraction of simulated trajectories from the initial vertex ending in fixation.
For a uniformly random initial vertex, the last wild type occupies a leaf with probability $0.957$, $0.974$, and $0.994$ at $r=1.5$, $2$, and $5$, respectively (\Cref{fig:charles}).
Every vertex with temperature above $1$ has an estimated fixation location probability below $0.012$.
At $r=1.5$, $2$, and $5$, the two likeliest fixation locations are the two leaves farthest from the root.
The two leaves have the same downstream neighbor and therefore equal fixation location probabilities.
At the expected fixation time, the best vertices to sample are the two likeliest fixation locations at $r=1.5$, $2$, and $5$.
At $r=1$, we simulated from each initial vertex until $4{,}000$ fixation events or $10^6$ trajectories, because a mutant near the root rarely reaches fixation.
From each of $24$ initial vertices, none of the $10^6$ simulated trajectories reached fixation.
By the exact neutral fixation probabilities~\cite{allen2015molecular}, the initial vertex is one of the $24$ vertices with probability below $10^{-5}$, conditional on fixation.
Omitting the $24$ initial vertices therefore changes every fixation location probability by less than $10^{-5}$.
For a uniformly random initial vertex, the last wild type occupies a leaf with probability $0.956$ at $r=1$.
The likeliest fixation location at $r=1$ is the leaf adjacent to the root, with fixation location probability $0.365$.
At the expected fixation time, the best vertices to sample are the third and fourth likeliest fixation locations at $r=1$, two leaves with the same downstream neighbor.

\begin{proposition}[Final phase of fixation, Lean \Cref{lean:factorization}]\label{prop:endgame}
Let $G$ be a directed graph without self-loops on which the process is absorbed with probability one from every mutant set.
Suppose every vertex has at least one outgoing edge and $\theta_v>0$, and suppose $\rho_r^G(u)>0$.
Conditional on fixation from the initial vertex $u$, the expected number of time steps with exactly one wild type left is
\[
T^{\mathrm{end}}(u)=\frac{r(N-1)+1}{r}\sum_{v\in V}L_r^G(v\mid u)\,\frac{\rho(V\setminus\{v\})}{\theta_v}.
\]
\end{proposition}

\begin{proof}
Before each active step from the mutant set $X$, the process spends a geometric number of time steps with mean $W(X)/a(X)$.
Here $W(X)$ is the total fitness and $a(X)$ is the total weight of active steps (\Cref{sec:computation}).
The holding time and the following active step are independent.
Conditional on fixation, the expected number of active steps from the mutant set $X$ is $n_u(X)\,\rho(X)/\rho_r^G(u)$.
The expected number of time steps with mutant set $X$ is therefore $n_u(X)\,\rho(X)\,W(X)/\bigl(a(X)\,\rho_r^G(u)\bigr)$.
For $X=V\setminus\{v\}$ we have $W(X)=r(N-1)+1$ and $a(X)=r\theta_v+1$, because $v$ is replaced at rate $r\theta_v$ and every reproduction of $v$ replaces a mutant.
\Cref{prop:factorization} implies $n_u(V\setminus\{v\})=\rho_r^G(u)\,L_r^G(v\mid u)\,(r\theta_v+1)/(r\theta_v)$.
Substituting and summing over $v$ proves the proposition.
\end{proof}

Since $\rho(V\setminus\{v\})\geq r\theta_v/(r\theta_v+1)$, the fixation time satisfies $T_r^G(u)\geq T^{\mathrm{end}}(u)\geq (r(N-1)+1)\sum_v L_r^G(v\mid u)/(r\theta_v+1)$.
Upper bounding every $\rho(V\setminus\{v\})$ with $1$ in \Cref{prop:endgame} yields the upper bound $T^{\mathrm{end}}(u)\leq\frac{r(N-1)+1}{r}\sum_v L_r^G(v\mid u)/\theta_v$.
The upper bound depends only on the fixation location probabilities and the temperatures.
On a regular graph every vertex has temperature $1$, so $T^{\mathrm{end}}(u)$ is at most $N$ time steps.
Consider all $11{,}117$ connected graphs with $N=8$ vertices at $r=2$, with a uniformly random initial vertex.
On the $11{,}117$ graphs, $T^{\text{end}}$ is within $2.5\%$ of the previously calculated general graph upper bound.

A graph is an amplifier of selection at $r>1$ if the fixation probability for a uniformly random initial vertex
exceeds the fixation probability of a well-mixed population with $N$ individuals.
The amplification is the ratio of the two fixation probabilities.
Consider the complete bipartite graph with parts of $m$ and $100-m$ vertices at $r=2$.
A vertex in the part with $m$ vertices has temperature $(100-m)/m$.
A vertex in the part with $100-m$ vertices has temperature $m/(100-m)$.
\Cref{tab:amplifiers} lists exact values for four values of $m$.
As $m$ decreases from $50$ to $1$, the amplification, the mean inverse temperature of the fixation location, $T^{\text{end}}$, and the fixation time increase.
We see that $T^{\text{end}}$ equals $99.5$ times the mean inverse temperature of the fixation location to within $0.1\%$, as \Cref{prop:endgame} predicts with $\rho(V\setminus\{v\})$ close to $1$.
On the connected undirected graphs with $N=5$ to $9$ vertices at $r=1.25$, $2$, and $10$, the fixation location of every amplifier except one has a larger mean inverse temperature than a uniformly random vertex.
The long $T^{\text{end}}$ is one source of the long fixation times of amplifiers \cite{tkadlec2019population}.

\begin{table}[h]
\centering\small
\caption{Amplification and $T^{\text{end}}$ on the complete bipartite graph with parts of $m$ and $100-m$ vertices at $r=2$, computed exactly. The initial vertex is uniformly random. The mean inverse temperature, $T^{\text{end}}$, and the fixation time are conditional on fixation. The units of $T^{\text{end}}$ and the fixation time ($T$) are in time steps.}
\label{tab:amplifiers}
\begin{tabular}{rrrrr}
\toprule
$m$ & Amplification & Mean $1/\theta_v$ of the fixation location & $T^{\text{end}}$ & Fixation time, $T$ \\
\midrule
$50$ & $1.00$ & $1.0$ & $99.5$ & $1{,}335$ \\
$20$ & $1.17$ & $3.6$ & $362$ & $3{,}186$ \\
$10$ & $1.31$ & $8.6$ & $856$ & $7{,}001$ \\
$1$ & $1.48$ & $99$ & $9{,}850$ & $77{,}269$ \\
\bottomrule
\end{tabular}
\end{table}

\section{Possible locations of fixation}
We consider the following decision problem.
There is only one initial mutant.
\begin{problem}[Fixability]
    \label{problem:fixability}
   \textit{%
   Given a directed graph $G=(V, E)$ and $r\geq 1$, the initial mutant location $s\in V$, and a target location $t\in V$,
   determine whether fixation of the Birth-death process at $t$ has positive probability.%
   }
\end{problem}
\noindent We characterize fixability exactly as reachability in the configuration graph.

\begin{definition}[Configuration graph]
For a directed graph $G=(V,E)$, let $\mathcal C(G)$ be the directed graph with vertex set $2^V$.
A state $X\subseteq V$ means the mutant set is $X$.
There is a directed edge $X\to X'$ in $\mathcal C(G)$ iff there exist $u,v\in V$ with $u\to v\in E$,
$\mathbf 1_X(u)\neq \mathbf 1_X(v)$, and $X'=X\triangle\{v\}$ (symmetric difference).
\end{definition}

\begin{theorem}[Exact characterization, all $s,t$, Lean \Cref{lean:finite}]
\label{thm:fixability-exact}
Let $G=(V,E)$ be finite, $r\geq 1$, and let $s,t\in V$ (possibly $s=t$).
Start with a single mutant at $s$, i.e.\ state $X_0=\{s\}$.
Write $X_t^\star:=V\setminus\{t\}$.
Then fixation at location $t$ has positive probability if and only if
\begin{enumerate}
\item $X_t^\star$ is reachable from $X_0$ in $\mathcal C(G)$, and
\item $X_t^\star\to V$ is an edge of $\mathcal C(G)$.
\end{enumerate}
Equivalently, the second condition is $\Gamma^-(t)\setminus\{t\}\neq\emptyset$
where $\Gamma^-\colon V\to 2^V$ maps a vertex to the set of incoming neighbors of the vertex.
\end{theorem}

\begin{proof}
Since $r\geq1$, every legal Birth-death event has strictly positive probability.
Hence an event has positive probability iff at least one finite legal event sequence realizes the event.
The event ``fixation occurs at location $t$'' means the mutant set is $X_t^\star$ at some time (all vertices mutant except $t$).
The next event then makes $t$ mutant, yielding state $V$.
In the configuration graph, the event is exactly a directed path
$X_0\rightsquigarrow X_t^\star\to V$.
So the two stated conditions are equivalent to fixability at $t$.
\end{proof}

\begin{corollary}[Basic impossibility checks for $s\neq t$, Lean \Cref{lean:finite}]
\label{cor:fixability-impossible-basic}
Assume $s\neq t$.
Fixation at location $t$ is impossible if either
\begin{enumerate}
\item there exists $v\in V$ with no directed path from $s$ to $v$, or
\item $\Gamma^-(t)\setminus\{t\}=\emptyset$.
\end{enumerate}
\end{corollary}

\begin{proof}
If some $v$ is unreachable from $s$, then $v$ can never become mutant.
Fixation is then impossible.
If $\Gamma^-(t)\setminus\{t\}=\emptyset$, then condition (2) of \Cref{thm:fixability-exact} fails.
Fixation at $t$ is then impossible.
\end{proof}

\noindent
Next, we derive conditions for fixability without examining the configuration graph.

\begin{definition}[Refillable path]
\label{def:refillable}
Assume $s\neq t$.
A simple directed path $P:u_0\to u_1\to\cdots\to u_k=t$ with $k\geq1$ is \emph{refillable (from $s$)} if the following three conditions hold.
\begin{enumerate}
\item $u_0\neq s$,
\item there exists $w\in \Gamma^-(u_0)\setminus\{u_0,u_1,\ldots,u_k\}$,
\item for every $v\in V\setminus\{t,u_0\}$, there exists a directed path from $s$ to $v$ in $G\setminus\{u_0\}$.
\end{enumerate}
\end{definition}

\begin{theorem}[Refillability implies fixability, Lean \Cref{lean:finite}]
\label{thm:refillable-sufficient}
Assume $s\neq t$. If there exists a refillable path from $s$ to $t$, then fixation at location $t$ is possible.
\end{theorem}

\begin{proof}
Fix a refillable path
$P:u_0\to u_1\to\cdots\to u_k=t$ with $k\geq 1$,
and let $w$ be a witness from \Cref{def:refillable}.
Since $u_0\neq s$, initially $u_0$ is wild type.
By condition (3), for each $v\in V\setminus\{t,u_0\}$ there is a directed path
from $s$ to $v$ in $G\setminus\{u_0\}$.
Legal Birth-death events along the paths make every vertex in
$V\setminus\{t,u_0\}$ mutant without a change at $u_0$.
Then $u_0$ is wild type.
Every vertex outside $\{t,u_0\}$ is mutant.
The vertex $t$ is mutant or wild type.

If $t$ is mutant, move the wild type from $u_0$ along the whole path $P$ as follows.
For $i=0,1,\ldots,k-1$, copy the type at $u_i$ to $u_{i+1}$.
Immediately after the copy, refill $u_i$ to mutant.
The refill uses $w\to u_0$ when $i=0$ and $u_{i-1}\to u_i$ when $i\geq 1$.
After the $k$ copies, $t$ is the unique wild-type vertex.
The process is then in state
$X_t^\star=V\setminus\{t\}$.

If $t$ is wild type, move the wild type from $u_0$ only as far as $u_{k-1}$,
refilling each previous vertex as above.
Then refill $u_{k-1}$ to mutant.
The refill uses $w\to u_0$ if $k=1$ and $u_{k-2}\to u_{k-1}$ if $k\geq 2$.
Again the resulting state is $X_t^\star=V\setminus\{t\}$.

In each case the final event $u_{k-1}\to t$ is legal and reaches the all-mutant
state $V$.
Hence fixation at location $t$ is possible.
\end{proof}

\begin{remark}
The absence of a refillable path
is \emph{not} by itself an impossibility criterion in general.
\smallskip

A concrete counterexample has $V=\{0,1,2,3,4\}$, $s=1$, $t=2$, and
\begin{equation}
E=\{0\to2,\ 0\to4,\ 1\to0,\ 2\to3,\ 3\to2,\ 4\to0\}.
\end{equation}

\begin{center}
\begin{tikzpicture}[
    >=Latex,
    vertex/.style={circle, draw, minimum size=8mm, inner sep=0pt, font=\small},
    special/.style={circle, draw, minimum size=8mm, inner sep=0pt, font=\small, fill=gray!15},
    edge/.style={->, thick}
]
    \node[special] (one) at (0,0) {$s=1$};
    \node[vertex]  (zero) at (2,0) {$0$};
    \node[special] (two) at (4,0) {$t=2$};
    \node[vertex]  (three) at (4,-1.8) {$3$};
    \node[vertex]  (four) at (2,-1.8) {$4$};

    \draw[edge] (one) -- (zero);
    \draw[edge] (zero) -- (two);

    \draw[edge, bend left=20] (two) to (three);
    \draw[edge, bend left=20] (three) to (two);

    \draw[edge, bend left=18] (zero) to (four);
    \draw[edge, bend left=18] (four) to (zero);
\end{tikzpicture}
\end{center}
There is no refillable path from $s$ to $t$.
Condition 1 of \Cref{def:refillable} requires $u_0\neq 1$.
The candidates for $u_0$ are $0$, $3$, and $4$.
For $u_0=0$, condition 3 fails, because every path from $1$ to $4$ uses $0$.
For $u_0=3$, condition 2 fails, because the only incoming neighbor of $3$ is $t=2$.
For $u_0=4$, the only candidate refillable path is $4\to 0\to 2$.
Condition 2 fails again, because $u_1=0$ is the only incoming neighbor of $4$.

Fixation at $t=2$ is nevertheless possible, for instance via
$\{1\}\to\{0,1\}\to\{0,1,2\}\to\{0,1,2,3\}\to\{1,2,3\}\to\{1,3\}\to\{0,1,3\}\to\{0,1,3,4\}$,
followed by one final step using $0\to2$.

\end{remark}

\subsection{Unidirectional bottlenecks}
\begin{figure}
  \centering
  \includegraphics[width=0.5\textwidth]{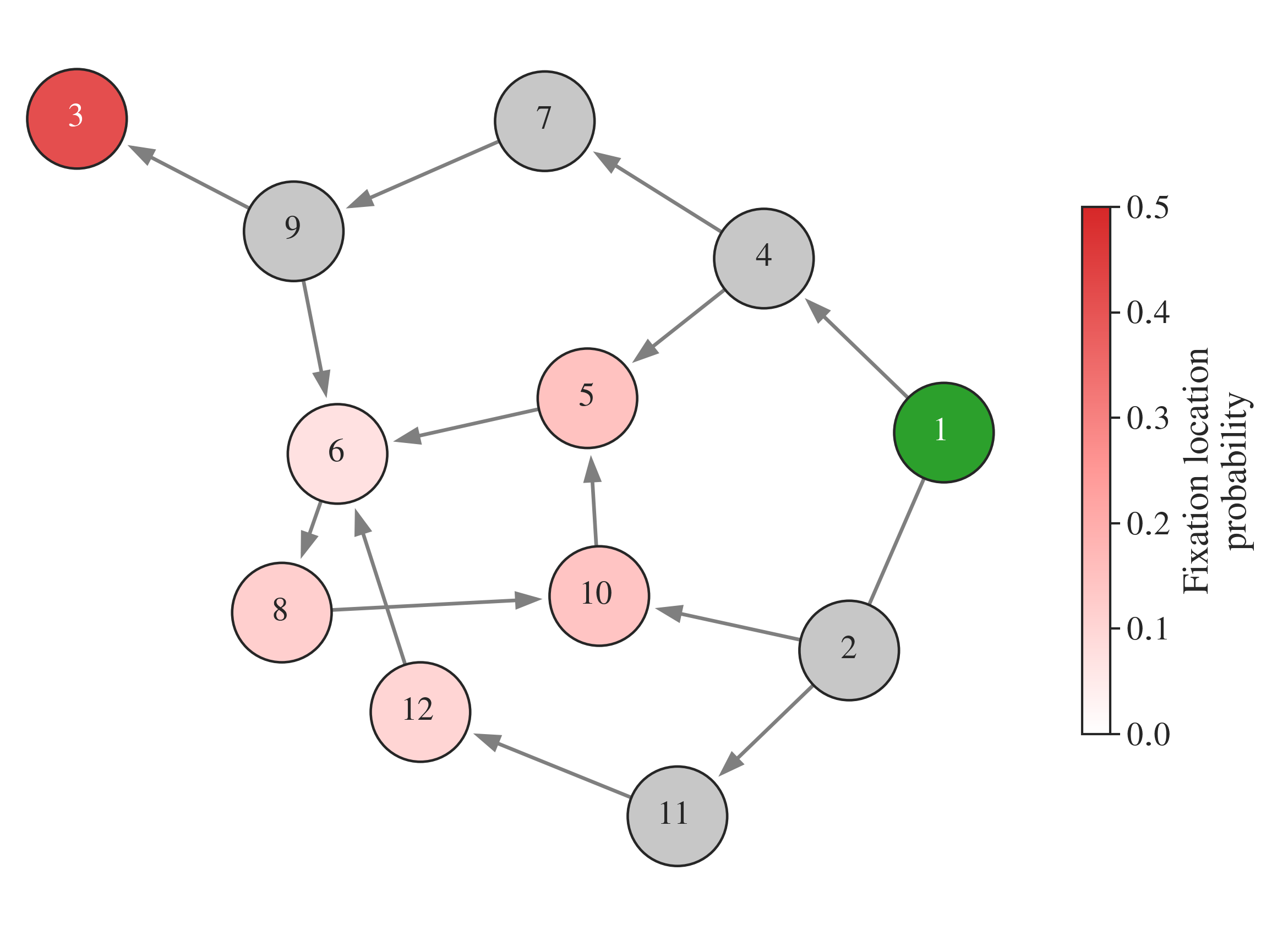}
  \caption{Some vertices of a graph with directed edges are impossible fixation locations at finite $r$. The graph has $N=12$ vertices. Arrows are directed edges. The line between vertices $1$ and $2$ is a bidirectional edge. A mutant with $r=1.25$ arises at vertex $1$ (green). Vertex shade is the fixation location probability of the vertex. Gray vertices and the initial vertex are impossible fixation locations. Vertex $3$ has temperature $1/2$ and fixation location probability $0.41$. Vertex $6$ has temperature $5/2$ and fixation location probability $0.07$.}
  \label{fig:example12}
  \figtoc{Some vertices of a graph with directed edges are impossible fixation locations at finite $r$}
\end{figure}

The impossible fixation locations of \Cref{fig:example12} satisfy the following sufficient condition.

\begin{proposition}[Unidirectional bottlenecks, Lean \Cref{lean:finite}]\label{prop:gateway}
Let $u$ be the initial mutant location and let $t\neq u$.
Suppose
(i) some vertex $w\neq t$ can be reached from $u$ only by directed paths through $t$, and
(ii) in every trajectory ending in fixation, once a mutant occupies $t$, a mutant occupies $t$ at every later time.
Then $t$ is not a possible fixation location.
If (ii) holds for $t=u$, then $u$ is not a possible fixation location either.
\end{proposition}

\begin{proof}
Consider a trajectory ending in fixation.
The mutant at $w$ at the end descends from the initial mutant through offspring placements along a directed path from $u$ to $w$.
By (i) the path from $u$ to $w$ passes through $t$.
A mutant therefore occupies $t$ at some time before the end.
By (ii) a mutant occupies $t$ from then on.
The last step of the trajectory replaces the last wild type.
The last wild type is therefore not at $t$.
For $t=u$, a mutant occupies $u$ at the start and, by (ii), at every later time.
\end{proof}

\begin{lemma}[Single incoming neighbor, Lean \Cref{lean:finite}]\label{lem:single-in}
Let $t\neq u$.
If $t$ has exactly one incoming neighbor $x$ and condition (ii) holds for $x$, then condition (ii) holds for $t$.
\end{lemma}

\begin{proof}
Vertex $t$ receives offspring only from $x$.
When a mutant first occupies $t$, the mutant is the offspring of a mutant at $x$.
By (ii) a mutant occupies $x$ afterwards.
So $t$ never again receives a wild-type offspring.
\end{proof}

In \Cref{fig:example12}, vertices 1 and 2 receive offspring only from each other.
Once vertices 1 and 2 are wild type at the same time, no mutant can enter vertex 1 or vertex 2 again.
Fixation is then impossible.
A wild type can occupy vertex 1 again only as the offspring of a wild type at vertex 2.
Condition (ii) therefore holds for vertex 1.
By \Cref{lem:single-in}, condition (ii) holds for vertex 2 and for vertices 4, 7, and 9 along the chain $1\to4\to7\to9$.
Condition (ii) also holds for vertex 11, whose only incoming neighbor is vertex 2.
Each of the six vertices is the only entrance to some other vertex.
Condition (i) therefore holds.
By \Cref{prop:gateway} the six vertices are impossible fixation locations, in agreement with the exact computation.

\subsection{Colonization on directed graphs}

For a directed graph $H$ and a vertex $u$ of $H$, let $R_H(u)$ be the set of vertices reachable from $u$ by directed paths in $H$, including $u$.
For a vertex $t$, let $G\setminus\{t\}$ be the graph obtained from $G$ by deleting $t$ and every edge at $t$.

\begin{proposition}[Possible fixation locations under colonization, Lean \Cref{lean:col}]\label{prop:colonization-directed}
Let $G=(V,E)$ be a finite directed graph and let $u,t\in V$.
For the colonization process ($r\to\infty$) started from a single mutant at $u$, the vertex $t$ is a possible fixation location if and only if
\[
t\neq u,\qquad R_{G\setminus\{t\}}(u)=V\setminus\{t\},\qquad \Gamma^-(t)\setminus\{t\}\neq\emptyset .
\]
\end{proposition}

\begin{proof}
Under colonization only mutants reproduce.
A vertex occupied by a mutant therefore remains occupied by a mutant.
The mutant set grows by one vertex at each active step.

Suppose $t$ is the fixation location with positive probability.
The initial vertex is occupied by a mutant at every time, so $t\neq u$.
Before the last active step the mutant set is $V\setminus\{t\}$.
A mutant never occupied $t$ before the last active step, because a mutant at $t$ is never replaced.
Each vertex $w\neq t$ received a mutant offspring from an in-neighbor occupied by a mutant.
Following the parents back from $w$ yields a directed path from $u$ to $w$ through vertices of $V\setminus\{t\}$.
Hence $R_{G\setminus\{t\}}(u)=V\setminus\{t\}$.
The last active step places a mutant offspring at $t$ from an in-neighbor of $t$, so $\Gamma^-(t)\setminus\{t\}\neq\emptyset$.

Conversely, suppose the three conditions hold.
Order the vertices of $V\setminus\{t\}$ by a breadth-first search from $u$ in $G\setminus\{t\}$.
Each vertex after $u$ in the order has an in-neighbor earlier in the order.
At each active step, the event placing a mutant offspring at the next vertex in the order has positive probability.
The mutant set therefore reaches $V\setminus\{t\}$ with positive probability.
In the mutant set $V\setminus\{t\}$, every in-neighbor of $t$ other than $t$ is occupied by a mutant.
The next active step places a mutant offspring at $t$.
Hence $t$ is the fixation location with positive probability.
\end{proof}

Under colonization condition (ii) of \Cref{prop:gateway} holds for every vertex.
The impossibility half of \Cref{prop:colonization-directed} therefore also follows from \Cref{prop:gateway}.
On a connected undirected graph and for $t\neq u$, the condition $R_{G\setminus\{t\}}(u)=V\setminus\{t\}$ holds if and only if $t$ is not a cut vertex.
\Cref{prop:colonization} below is therefore the special case of \Cref{prop:colonization-directed} for undirected graphs.
The three conditions require one breadth-first search in $G\setminus\{t\}$ for each vertex $t$.
The possible fixation locations of the colonization process are therefore computed in $O\bigl(N(N+|E|)\bigr)$ time on every directed graph.
An exhaustive search over mutant sets agrees with \Cref{prop:colonization-directed} on every directed graph with $3$ and $4$ vertices, from every initial vertex, and on $2{,}000$ random directed graphs with $5$ to $7$ vertices.

\subsection{Colonization on undirected graphs}

\begin{proposition}[Colonization on undirected graphs, Lean \Cref{lean:col}]\label{prop:colonization}
Let $G$ be a connected undirected graph with $N\geq 2$ vertices.
Consider the colonization process ($r\to\infty$) started from a single mutant at $u$.
A vertex $t$ is a possible fixation location if and only if $t\neq u$ and $t$ is not a cut vertex of $G$.
\end{proposition}

\begin{proof}
Under colonization only mutants reproduce.
A vertex occupied by a mutant therefore remains occupied by a mutant.
The mutant set grows by one neighboring vertex at a time.
The initial vertex is occupied by a mutant throughout.
Hence $t=u$ is impossible.
If $t\neq u$ is a cut vertex, some connected component $K$ of $G\setminus\{t\}$ does not contain $u$.
Every path from $u$ to $K$ passes through $t$.
So $t$ becomes mutant before any vertex of $K$.
Hence $t$ is not last.
If $t\neq u$ is not a cut vertex, $G\setminus\{t\}$ is connected.
The mutant lineage can then occupy $G\setminus\{t\}$ one neighboring vertex at a time, with positive probability at each step.
Finally the lineage occupies $t$, a vertex with at least one neighbor.
Hence $t$ is the fixation location with positive probability.
\end{proof}

For all connected undirected graphs with $N=5$ vertices,
we compute the fixation locations of the colonization process (see Figure~3 of the main text).
The initial mutant location cannot be a fixation location.
A cut vertex, a vertex whose removal disconnects the graph, cannot be a fixation location. Every vertex other than the initial mutant location and the cut vertices can be the fixation location, on any connected undirected graph. Under colonization a mutant is never replaced by a wild type. The mutant lineage must therefore pass through a cut vertex before the part of the graph beyond the cut vertex. If a vertex is not a cut vertex, the mutant lineage can occupy the rest of the graph before the vertex (\Cref{prop:colonization}).

Vertex $v$ is a possible fixation location from initial vertex $u$ if and only if the mutant set $V\setminus\{v\}$ is reachable from the mutant set $\{u\}$ by transitions of positive probability.
At finite $r\geq 1$ the transitions are the gain and the loss of one mutant.
Reachability therefore does not depend on the value of $r$.
We checked reachability on all $986$ connected undirected graphs with $N=5$, $6$, and $7$ vertices.
At finite $r$ only the three paths and the three stars have impossible fixation locations.
On the other $980$ graphs every vertex is a possible fixation location from every initial vertex.
Under colonization the impossible fixation locations are the initial vertex and the cut vertices on all $6{,}748$ pairs of a graph and an initial vertex, in agreement with \Cref{prop:colonization}.

\section{Directed cycles}\label{sec:directed}

A \emph{directed cycle}, $D_N$, is a chain of $N$ vertices $v_1,\ldots,v_N$ with a directed edge from each vertex to the next
(see Figure~4\textbf{a} of the main text).
The last vertex has a directed edge to the first.
When $r \to \infty$, the fixation location is the vertex with an edge directed into the initial mutant location.
In neutral evolution, each location is roughly equally likely to be the fixation location in large population sizes (see Figure~4\textbf{b} of the main text).
For $r>1$ the fixation location depends on the initial mutant location
(see Figure~4\textbf{c,d} of the main text).

Here the model is a cycle on $N$ vertices oriented in the order $1\to 2 \to \dots \to N \to 1$.
The initial mutant is at vertex 1.
Let $p_k$ denote the conditional probability of fixation happening at vertex $k$, conditional on the mutant being the fixating type. 
A time step of the process is \emph{active} if the number of mutants changes in the step.

\label{par:block}%
Let $r\ge1$.
At every time step the mutants form a block of adjacent vertices.
Each offspring is placed at the next vertex of the cycle.
The block therefore grows only at the head and shrinks only at the tail.
The head advances when the mutant at the head places an offspring at the next vertex.
The tail retreats when the wild type behind the tail places an offspring at the tail.
At an active step the head advances with probability $\alpha\coloneqq r/(1+r)$.
At an active step the tail retreats with probability $\beta\coloneqq 1/(1+r)$.
The length of the block therefore performs a random walk on $\{0,1,\ldots,N\}$ started at $1$.
The walk moves up with probability $\alpha$ and down with probability $\beta$.
Fixation occurs when the walk reaches $N$ before $0$.
The fixation probability $\rho$ equals $(1-r^{-1})/(1-r^{-N})$ for $r>1$ and $1/N$ for $r=1$.

Let $H$ be the number of head advances until fixation, conditional on fixation.
The last head advance places a mutant at the vertex of the last wild type.
The fixation location is therefore vertex $1+H$, with vertex labels taken modulo $N$.
Let $K$ be the number of tail retreats until fixation.
The block grows from length $1$ to length $N$.
Hence $H=K+N-1$.
The walk therefore takes $n=H+K=2H-N+1$ active steps.

\begin{lemma}[Exact law of the fixation location, Lean \Cref{lean:gf,lean:fourier}]\label{lem:cycle-pgf}
Let $P_N(y)\coloneqq\sum_{m\ge0}\binom{N}{2m+1}y^{m}$.
For every complex $z$ with $|z|\le1$,
\begin{equation}\label{eq:cycle-pgf}
\mathbb{E}\bigl[z^{H}\bigr]=\frac{(2\alpha z)^{N-1}}{\rho\,P_N(1-4\alpha\beta z)}.
\end{equation}
For $k=1,2,\ldots,N$, the fixation location is vertex $k$ with probability
\begin{equation}\label{eq:cycle-fourier}
p_k=\frac{1}{N}\sum_{\ell=0}^{N-1}e^{-2\pi i\ell(k-1)/N}\,\mathbb{E}\bigl[e^{2\pi i\ell H/N}\bigr].
\end{equation}
\end{lemma}

\begin{proof}[Proof of \Cref{lem:cycle-pgf}]
Consider the walk with absorbing states $0$ and $N$.
For $x\in\{0,1,\ldots,N\}$, let $\mathbb{E}_x$ denote the expectation over the trajectories of the walk started at position $x$.
Let $n$ be the number of steps until the walk first reaches $0$ or $N$.
The number $n$ is a random variable.
The distribution of $n$ under $\mathbb{E}_x$ depends on the start $x$.
Each step of the walk is one active step of the population.
For $0<s<1$, let
\begin{equation}
G_x(s)\coloneqq\mathbb{E}_x\bigl[s^{n}\,\mathbbm{1}\{\text{the walk reaches }N\text{ before }0\}\bigr].
\end{equation}
A walk started at $0$ never reaches $N$.
Hence $G_0(s)=0$.
A walk started at $N$ takes $n=0$ steps.
Hence $G_N(s)=1$.
For $0<x<N$, we condition on the first step of the walk.
The first step goes to position $x+1$ with probability $\alpha$ and to position $x-1$ with probability $\beta$.
After the first step, the walk continues as a walk started at the new position.
By the Markov property, the continuation is independent of the first step.
The continuation takes $n-1$ steps.
The continuation reaches $N$ before $0$ exactly when the whole walk reaches $N$ before $0$.
The first step contributes the factor $s$ to $s^{n}$.
This first step analysis yields
\begin{equation}
G_x(s)=\alpha s\,G_{x+1}(s)+\beta s\,G_{x-1}(s),\qquad 0<x<N.
\end{equation}
The equation $\alpha s\lambda^{2}-\lambda+\beta s=0$ has the roots $\lambda_{\pm}\coloneqq(1\pm w)/(2\alpha s)$, with $w\coloneqq\sqrt{1-4\alpha\beta s^{2}}$.
For $0<s<1$, $w>0$ and the roots $\lambda_{+}$ and $\lambda_{-}$ are distinct.
Every solution of the recurrence is therefore $A\lambda_{+}^{x}+B\lambda_{-}^{x}$ for constants $A$ and $B$.
The boundary values $G_0(s)=0$ and $G_N(s)=1$ determine $G_x(s)=(\lambda_{+}^{x}-\lambda_{-}^{x})/(\lambda_{+}^{N}-\lambda_{-}^{N})$.
The binomial theorem yields $\lambda_{+}^{N}-\lambda_{-}^{N}=2w\,P_N(w^{2})/(2\alpha s)^{N}$ and $\lambda_{+}-\lambda_{-}=w/(\alpha s)$.
Therefore
\begin{equation}
G_1(s)=\frac{(2\alpha s)^{N-1}}{P_N(1-4\alpha\beta s^{2})}.
\end{equation}
Every fixation trajectory has $n=2H-N+1$ active steps.
Hence the expectation of $z^{H}$ on the event of fixation equals $z^{(N-1)/2}G_1(\sqrt{z})=(2\alpha z)^{N-1}/P_N(1-4\alpha\beta z)$ for $0<z<1$.
Division by the fixation probability $\rho$ yields Eq.~\eqref{eq:cycle-pgf} for $0<z<1$.
The left side of Eq.~\eqref{eq:cycle-pgf} is a power series with nonnegative coefficients summing to $1$.
The right side is a rational function of $z$.
By the identity theorem the two sides agree on the open unit disk.
The left side is bounded by $1$ on the closed unit disk.
The right side therefore has no pole on the closed unit disk.
By continuity, the two sides agree on the closed unit disk.
Eq.~\eqref{eq:cycle-fourier} follows from the orthogonality of the $N^\text{th}$ roots of unity.
The sum $\frac1N\sum_{\ell=0}^{N-1}e^{2\pi i\ell(h-k+1)/N}$ equals $1$ if $1+h\equiv k\pmod N$ and $0$ otherwise.
\end{proof}

\subsection{Neutral evolution}
\begin{proposition}[Neutral evolution, Lean \Cref{lean:neutral}]\label{prop:cycle-neutral}
If $r=1$ then $p_k=1/N+O\bigl(N^{-1/2}e^{-\sqrt{N/2}}\bigr)$, with an implied constant independent of $k$.
\end{proposition}

By exact computation, the largest deviation $\max_k|Np_k-1|$ equals $0.21$, $2.1\times10^{-2}$, $1.0\times10^{-3}$, $1.3\times10^{-5}$, and $2.5\times10^{-8}$ for $N=10$, $20$, $40$, $80$, and $160$.
The terms $\ell=1$ and $\ell=N-1$ of Eq.~\eqref{eq:cycle-fourier} dominate the deviation.

\begin{proof}[Proof of \Cref{prop:cycle-neutral}]
At $r=1$, \Cref{lem:cycle-pgf} implies $\mathbb{E}[z^{H}]=Nz^{N-1}/P_N(1-z)$.
Let $\omega\coloneqq e^{2\pi i/N}$.
The term $\ell=0$ of Eq.~\eqref{eq:cycle-fourier} equals $1/N$.
The other terms contribute
\begin{equation}\label{eq:cycle-neutral-dev}
p_k-\frac{1}{N}=\sum_{\ell=1}^{N-1}\frac{\omega^{-\ell k}}{P_N(1-\omega^{\ell})}.
\end{equation}
Fix $\ell\in\{1,\ldots,N-1\}$.
The number $1-\omega^{\ell}$ equals $2\sin(\pi\ell/N)\,e^{i(\pi\ell/N-\pi/2)}$ and has argument in $(-\pi/2,\pi/2)$.
Let $x$ be the square root of $1-\omega^{\ell}$ with argument in $(-\pi/4,\pi/4)$.
Then $|x|^{2}=2\sin(\pi\ell/N)\le2$ and $\operatorname{Re}x\ge|x|/\sqrt2$.
Let $s\coloneqq|x|/\sqrt2=\sqrt{\sin(\pi\ell/N)}\le1$.
Hence $|1+x|\ge1+s$ and $|1-x|^{2}=1-2\operatorname{Re}x+|x|^{2}\le1-\sqrt2|x|+|x|^{2}\le1$.
The binomial theorem yields $P_N(x^{2})=\bigl((1+x)^{N}-(1-x)^{N}\bigr)/(2x)$.
Therefore
\begin{equation}
\bigl|P_N(1-\omega^{\ell})\bigr|\ge\frac{(1+s)^{N}-1}{2\sqrt2\,s}.
\end{equation}
The terms $\ell$ and $N-\ell$ have the same value of $s$.
We therefore take $\ell\le N/2$.
The inequality $\sin t\ge2t/\pi$ for $0\le t\le\pi/2$ implies $\sin(\pi\ell/N)\ge2\ell/N$.
Hence $s\ge\sqrt{2\ell/N}\ge2/N$.
The inequalities $(1+s)^{N}\ge1+Ns\ge2$ and $\log(1+s)\ge s/2$ imply $(1+s)^{N}-1\ge e^{Ns/2}/2$.
The function $s\mapsto s\,e^{-Ns/2}$ decreases for $s\ge2/N$.
Hence
\begin{equation}
\frac{1}{\bigl|P_N(1-\omega^{\ell})\bigr|}\le4\sqrt2\,s\,e^{-Ns/2}\le8\sqrt{\ell/N}\;e^{-\sqrt{\ell N/2}}.
\end{equation}
For $N\ge2$ and $\ell\ge1$, the exponent satisfies $\sqrt{\ell N/2}\ge\sqrt{N/2}+\sqrt{\ell}-1$.
Summation of the bound over $\ell$ in Eq.~\eqref{eq:cycle-neutral-dev} yields
\begin{equation}
\max_k\Bigl|p_k-\frac{1}{N}\Bigr|\le\frac{16}{\sqrt N}\sum_{\ell\ge1}\sqrt{\ell}\,e^{-\sqrt{\ell N/2}}\le\frac{16\,C}{\sqrt N}\,e^{-\sqrt{N/2}},
\end{equation}
with $C\coloneqq\sum_{\ell\ge1}\sqrt{\ell}\,e^{1-\sqrt{\ell}}<10.6$.
\end{proof}

\subsection{Constant selection}
We now fix $r>1$.

\begin{theorem}[Directed cycle under constant selection, Lean \Cref{lean:mean,lean:var,lean:clt}]\label{thm:cycle-selection}
Let $r>1$.
The number $H$ of head advances has mean
\begin{equation}\label{eq:cycle-mean}
\mathbb{E}[H]=(N-1)\frac{r}{r-1}-\frac{r+1}{(r-1)^{2}}+\frac{N(r+1)}{(r-1)(r^{N}-1)}
\end{equation}
and variance
\begin{equation}\label{eq:cycle-var}
\operatorname{Var}[H]=\frac{r(r+1)}{(r-1)^{3}}\Bigl(N-\frac{2(r+1)}{r-1}\Bigr)+O\bigl(N^{2}r^{-N}\bigr).
\end{equation}
As $N\to\infty$, the random variable $(H-\mathbb{E}[H])/\sqrt{\operatorname{Var}[H]}$ converges in distribution to a standard normal random variable.
\end{theorem}

\begin{corollary}[Concentration of the fixation location, Lean \Cref{lean:clt}]\label{cor:cycle-region}
Let $r>1$ and $a>0$.
Let $m_N\coloneqq(N-1)r/(r-1)-(r+1)/(r-1)^{2}$ and $\sigma_r^{2}\coloneqq r(r+1)/(r-1)^{3}$.
Let $\Phi$ be the standard normal distribution function.
The probability of a fixation location within $a\sqrt{N}$ vertices of vertex $1+m_N$ (modulo $N$) has limit inferior at least $2\Phi(a/\sigma_r)-1$ as $N\to\infty$.
\end{corollary}

\begin{figure}
  \centering
  \includegraphics[width=0.95\textwidth]{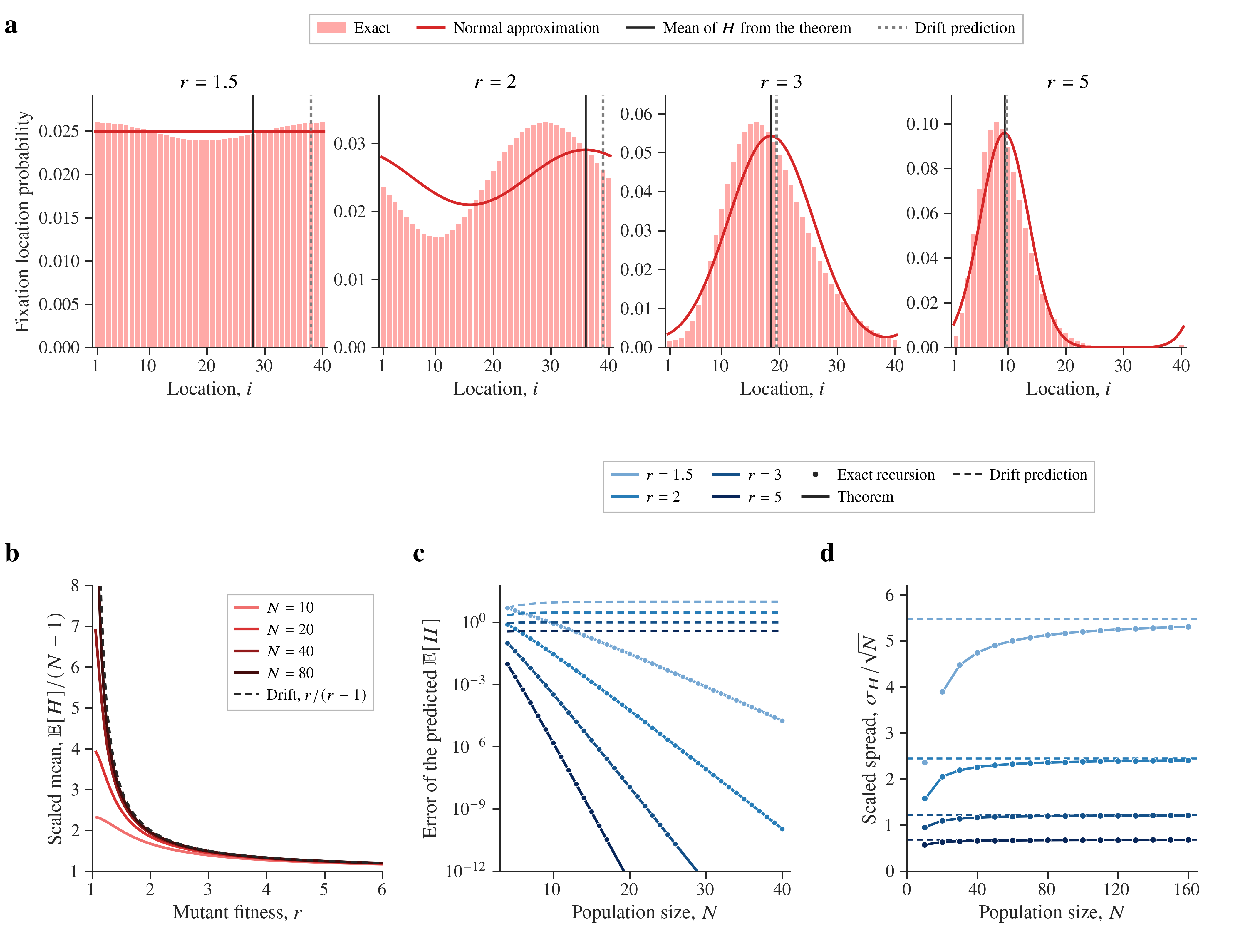}
  \caption{The fixation location on the directed cycle under constant selection agrees with \Cref{thm:cycle-selection}.
  The initial mutant is at vertex $1$.
  The number $H$ of head advances is counted until fixation, conditional on fixation.
  The fixation location is vertex $1+H$, with vertex labels taken modulo $N$.
  Exact values come from a recursion over the length of the block of mutants and the number of tail retreats.
  \textbf{a,} Fixation location probability at $N=40$ for $r=1.5$, $2$, $3$, and $5$.
  The solid line is at vertex $1+\mathbb{E}[H]$ with $\mathbb{E}[H]$ from \Cref{thm:cycle-selection}.
  The dotted line is at the drift prediction, vertex $1+(N-1)r/(r-1)$.
  The red curve is the normal density with the mean and the variance of \Cref{thm:cycle-selection}, wrapped around the cycle.
  At $N=40$ the distribution of $H$ is skewed.
  The most likely fixation location is therefore a few vertices before vertex $1+\mathbb{E}[H]$.
  For $r=1.5$ the standard deviation of $H$ is $30$, comparable with $N=40$.
  The fixation location then spreads over the whole cycle.
  \textbf{b,} Mean number of head advances divided by $N-1$ against the mutant fitness, for $N=10$, $20$, $40$, and $80$.
  The dashed curve is the drift value $r/(r-1)$.
  \textbf{c,} Error of two predictions of $\mathbb{E}[H]$ against the population size.
  Dashed lines are the error of the drift prediction $(N-1)r/(r-1)$.
  The error of the drift prediction tends to $(r+1)/(r-1)^{2}$.
  Solid lines are the error of $(N-1)r/(r-1)-(r+1)/(r-1)^{2}$.
  The error of the corrected prediction decays exponentially in $N$.
  Markers are the error computed with the exact recursion, drawn above $10^{-11}$.
  \textbf{d,} Standard deviation $\sigma_H$ of $H$ divided by $\sqrt{N}$ against the population size.
  Markers are exact.
  Solid curves are the variance of \Cref{thm:cycle-selection} without the exponentially small term.
  Dashed lines are the leading order $\sqrt{r(r+1)/(r-1)^{3}}$.}
  \label{fig:cycle-selection}
  \figtoc{The fixation location on the directed cycle under constant selection agrees with \protect\Cref{thm:cycle-selection}}
\end{figure}

The second term of Eq.~\eqref{eq:cycle-mean} is the effect of the conditioning on fixation.
Trajectories with an early return of the tail to the head end in extinction.
Conditioning on fixation removes the trajectories ending in extinction.
The remaining trajectories have fewer tail retreats.
The third term decays exponentially in $N$.
For $r=3$, vertex $1+m_N$ is vertex $N/2-3/2$ modulo $N$.
The fixation location therefore concentrates near the vertex opposite the initial mutant.
The spread $\sigma_r\sqrt{N}$ is small compared with $N$ when $(r-1)^{3}N$ is large compared with $r(r+1)$.
For smaller $r$, the fixation location spreads over the whole cycle, as in neutral evolution.
\Cref{fig:cycle-selection} compares \Cref{thm:cycle-selection} with an exact recursion over the length of the block and the number of tail retreats.

\begin{proof}[Proof of \Cref{thm:cycle-selection}]
For complex $\theta$ near $0$, let $y(\theta)\coloneqq1-4\alpha\beta e^{\theta}$, let $w(\theta)\coloneqq\sqrt{y(\theta)}$, and let $c(\theta)\coloneqq(1-w(\theta))/(1+w(\theta))$.
At $\theta=0$ the values are $w(0)=(r-1)/(r+1)$ and $c(0)=1/r$.
The binomial theorem yields $P_N(y)=(1+w)^{N}(1-c^{N})/(2w)$.
By \Cref{lem:cycle-pgf},
\begin{equation}\label{eq:cycle-cgf}
\log\mathbb{E}\bigl[e^{\theta H}\bigr]=A_N+(N-1)\theta-N\log\bigl(1+w(\theta)\bigr)+\log\bigl(2w(\theta)\bigr)-\log\bigl(1-c(\theta)^{N}\bigr),
\end{equation}
with $A_N\coloneqq(N-1)\log(2\alpha)-\log\rho$ independent of $\theta$.
The mean is the first derivative of Eq.~\eqref{eq:cycle-cgf} at $\theta=0$.
The first derivatives of the four terms depending on $\theta$ are $N-1$, $N/(r-1)$, $-2r/(r-1)^{2}$, and $N(r+1)/\bigl((r-1)(r^{N}-1)\bigr)$.
The sum is Eq.~\eqref{eq:cycle-mean}.
The variance is the second derivative of Eq.~\eqref{eq:cycle-cgf} at $\theta=0$.
The second derivatives of the last three terms are $Nr(r+1)/(r-1)^{3}$, $-2r(r+1)^{2}/(r-1)^{4}$, and $O(N^{2}r^{-N})$.
The sum is Eq.~\eqref{eq:cycle-var}.

For the central limit theorem, choose $\delta>0$ and $c_0<1$ with $|y(\theta)-y(0)|\le y(0)/2$ and $|c(\theta)|\le c_0$ for $|\theta|\le\delta$.
On the disk $|\theta|\le\delta$ the functions $\log(1+w(\theta))$ and $\log(2w(\theta))$ are analytic.
On the same disk the last term of Eq.~\eqref{eq:cycle-cgf} is $O(c_0^{N})$, uniformly in $\theta$.
Let $\sigma_N^{2}\coloneqq\operatorname{Var}[H]$ and fix a real number $t$.
By Eq.~\eqref{eq:cycle-var}, $\sigma_N^{2}/N$ tends to $r(r+1)/(r-1)^{3}>0$.
For large $N$ the point $\theta\coloneqq it/\sigma_N$ lies in the disk.
The third derivative of Eq.~\eqref{eq:cycle-cgf} is $O(N)$ on the disk.
A Taylor expansion of Eq.~\eqref{eq:cycle-cgf} to second order therefore yields
\begin{equation}
\log\mathbb{E}\Bigl[\exp\Bigl(it\,\frac{H-\mathbb{E}[H]}{\sigma_N}\Bigr)\Bigr]=-\frac{t^{2}}{2}+O\Bigl(\frac{N|t|^{3}}{\sigma_N^{3}}\Bigr)=-\frac{t^{2}}{2}+O\bigl(N^{-1/2}\bigr).
\end{equation}
The characteristic function of the standardized $H$ therefore converges to $e^{-t^{2}/2}$ for every real $t$.
By the continuity theorem for characteristic functions, the standardized $H$ converges in distribution to a standard normal random variable.
\end{proof}

\begin{proof}[Proof of \Cref{cor:cycle-region}]
By Eq.~\eqref{eq:cycle-mean}, $|\mathbb{E}[H]-m_N|\le N(r+1)/\bigl((r-1)(r^{N}-1)\bigr)$.
The bound tends to $0$.
By Eq.~\eqref{eq:cycle-var}, $\operatorname{Var}[H]/N$ tends to $\sigma_r^{2}$.
The event $|H-m_N|\le a\sqrt{N}$ places the fixation location $1+H$ within $a\sqrt{N}$ vertices of vertex $1+m_N$ (modulo $N$).
By \Cref{thm:cycle-selection}, the probability of the event $|H-m_N|\le a\sqrt{N}$ tends to $2\Phi(a/\sigma_r)-1$.
\end{proof}

\section{Testing a hypothesized population structure}\label{sec:structure}

We test a theorized population structure against an alternative structure.
We cannot see the whole population.
We monitor a set $M$ of vertices, from the time step when a single mutant arises at a uniformly random vertex.
We repeat the observation over many independent trajectories.
Figure~4\textbf{e} to \textbf{g} of the main text applies the test to two structures with $N=7$ vertices.

A wild type at a monitored vertex rules out fixation.
A mutant at every monitored vertex does not prove fixation, because wild types can remain at the other vertices.
We therefore wait until fixation has occurred with high probability.
Let $\tau$ be the time step of fixation on a fixation trajectory.
The fixation time $T$ is the expected value of $\tau$.
By Markov's inequality $1-f_t=P(\tau>t)\leq T/t$ for every $t$.
Hence $f_t\geq0.95$ for $t\geq20T$.
To test two structures, we compute the fixation time of each structure exactly or as an upper bound.
Waiting $20$ times the larger fixation time then guarantees $f_t\geq0.95$ under each structure.
On connected undirected graphs, an upper bound on $T$ polynomial in $N$ follows from the potential function of D\'iaz et al.\ \cite{diaz2014approximating}.
For strongly connected directed graphs the fixation time is bounded by a polynomial in $N$ and in $1/\min_u\rho(u)$ \cite{brewster2024fixation}.
A trajectory ending in extinction before the end of the waiting time leaves a wild type at every monitored vertex.
We discard every trajectory with a wild type at a monitored vertex at the end of the waiting time.

\begin{figure}
  \centering
  \includegraphics[width=0.62\textwidth]{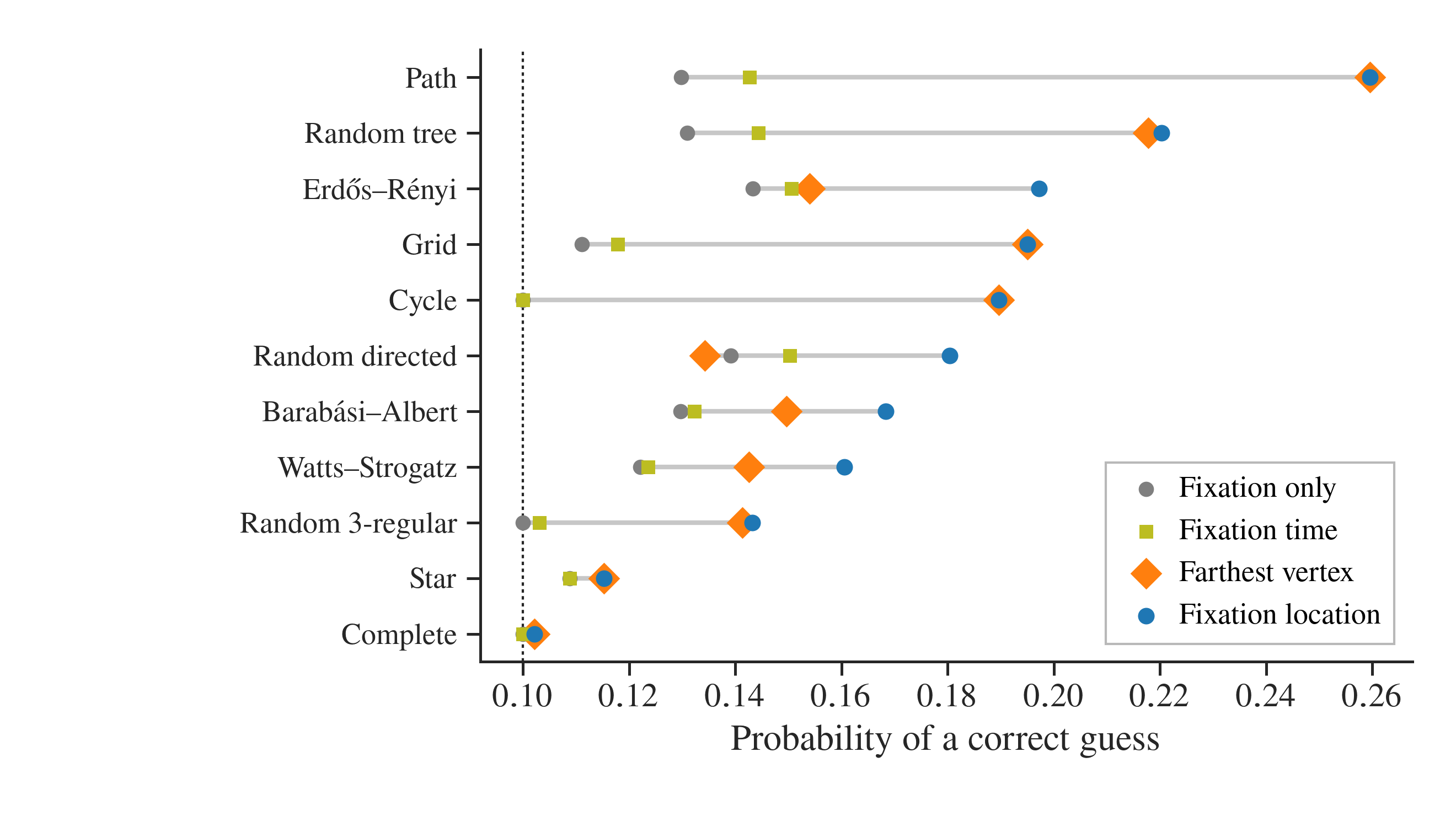}
  \caption{
    The fixation location provides information about the initial vertex.
    Probability of a correct guess of the initial vertex for graphs with $N=10$ vertices at $r=2$.
    The graphs are the cycle, path, star, complete graph, and $2\times5$ grid, and five instances each of six random graph families.
    The families are random $3$-regular graphs, Erd\H{o}s--R\'enyi graphs with mean degree $3.5$ conditioned on connectedness, Barab\'asi--Albert graphs with $m=2$, connected Watts--Strogatz graphs with $k=4$ and rewiring probability $0.2$, uniformly random labeled trees, and random directed graphs with two out-neighbors per vertex conditioned on strong connectedness.
    The probabilities are exact and averaged over $5$ instances of each random graph family.
    The initial vertex is uniformly random.
    The observer sees only trajectories ending in fixation.
    Before the observer sees the fixation location, the initial vertex is $u$ with probability $\rho_r^G(u)/\sum_w\rho_r^G(w)$.
    The best guess is the vertex with the largest probability of being the initial vertex, given the observation.
    Gray markers are the probability of a correct guess from fixation alone.
    Without further observation, the best guess is the vertex with the largest fixation probability.
    Olive markers are the probability of a correct guess from the fixation time.
    The fixation time distributions come from iterating the Birth-death chain until the probability of no absorption is below $10^{-11}$ for every initial vertex.
    Orange markers are the probability of a correct guess by the farthest vertex rule.
    Under the farthest vertex rule, the guess is the vertex farthest from the fixation location, with ties broken uniformly at random.
    Blue markers are the probability of a correct guess from the fixation location $v$, with the guess $u$ maximizing $\rho_r^G(u)\,L_r^G(v\mid u)$.
    The light gray line of each row spans the 4 markers.
    Markers with equal values are drawn on top of each other.
    The dotted line is $1/N$.
  }
  \label{fig:origin}
  \figtoc{The fixation location provides information about the initial vertex}
\end{figure}

The observation of one fixation event is the vertex of $M$ last occupied by a wild type.
If $M$ contains every vertex, the observation is the fixation location.
The probability of each observation under a given structure is computed exactly on an extended chain.
A state of the extended chain is a mutant set and a label.
The label is the last vertex of $M$ to be the only vertex of $M$ occupied by a wild type.
The mutant set follows the Birth-death chain conditioned on fixation.
At fixation the label is the observation.
The fixation location also provides information about the initial vertex (\Cref{fig:origin}).
The vertex farthest from the fixation location is not always the best guess of the initial vertex.

Let $P$ and $Q$ be the distributions of the observation under the true structure and under the other structure.
After $n$ fixation events we compute the probability of the $n$ observations under each structure.
The probability of the $n$ observations is the product of the probabilities of the individual observations.
We choose the structure with the larger probability.
For an observation with two values, the probability of a correct choice after $n$ fixation events is an exact binomial sum.
For an observation with more values, the probability of a correct choice comes from sequences of observations sampled from $P$.
Let $P^n(\mathbf{x})\coloneqq\prod_{i=1}^nP(x_i)$ and $Q^n(\mathbf{x})\coloneqq\prod_{i=1}^nQ(x_i)$ for a sequence $\mathbf{x}=(x_1,\ldots,x_n)$ of $n$ independent observations.
We choose wrong only on sequences with $Q^n(\mathbf{x})\geq P^n(\mathbf{x})$, with a tie counted as half.
On such a sequence $P^n(\mathbf{x})\leq\sqrt{P^n(\mathbf{x})\,Q^n(\mathbf{x})}$.
Summing over the sequences bounds the probability of a wrong choice by
\[
\sum_{\mathbf{x}}\sqrt{P^n(\mathbf{x})\,Q^n(\mathbf{x})}=\Bigl(\sum_x\sqrt{P(x)\,Q(x)}\Bigr)^n.
\]
The sum factorizes because $P^n$ and $Q^n$ are products over the $n$ independent observations.
By the Cauchy--Schwarz inequality, $B\coloneqq\sum_x\sqrt{P(x)\,Q(x)}$ satisfies $B\leq\sqrt{\sum_xP(x)}\,\sqrt{\sum_xQ(x)}=1$.
Equality holds only for $P=Q$.
Whenever $P\neq Q$, the probability of a wrong choice therefore decreases exponentially with $n$, at least as fast as $B^n$.
On a finite Markov chain, $1-f_t$ decreases exponentially in $t$, so waiting a larger multiple of the fixation time makes the error from unfinished trajectories small compared with $B^n$.

\section{Islands}\label{sec:islands}

Consider the Markov chain over the numbers of mutants on the two islands.
Fixation occurs through one of two states.
In the state with $n-1$ mutants on island $1$ and $n$ on island $2$, the last wild type is on island $1$.
In the state with $n$ and $n-1$ respectively, the last wild type is on island $2$.

For $m_{12}=10^{-3}$ and $m_{21}=10^{-1}$, the exact Markov chain gives the unbiased fitness $r^*\approx 1.26$, $1.30$, and $1.30$ for $N=20$, $40$, and $80$ (Figure~5\textbf{d} of the main text).

\subsection{On large islands under colonization the last wild type is on the lower temperature island}
In the colonization process only mutants reproduce.
In each time step a uniformly chosen mutant is the parent.
Let $u$ and $v$ be the numbers of wild types on islands $1$ and $2$, with $a\coloneqq n-u$ and $b\coloneqq n-v$ mutants.
Let
\begin{equation}
    \alpha\coloneqq(1-m_{12})a+m_{21}b,
    \qquad
    \beta\coloneqq m_{12}a+(1-m_{21})b.
\end{equation}
One time step replaces a wild type on island $1$ with probability $\alpha u/\bigl(n(a+b)\bigr)$ and a wild type on island $2$ with probability $\beta v/\bigl(n(a+b)\bigr)$.
The vertices of island $1$ have temperature $1-m_{12}+m_{21}$.
The vertices of island $2$ have temperature $1-m_{21}+m_{12}$.
Island $1$ has the lower temperature exactly when $m_{12}>m_{21}$.
Let $p_1^{(n)}$ be the probability of a fixation location on island $1$ for islands of $n$ individuals.

\begin{theorem}[Large islands under colonization, Lean \Cref{lean:large-islands}]\label{thm:large-islands}
Let $0<m_{21}<m_{12}\leq1$.
For the colonization process from a single mutant on island $1$ or on island $2$,
\[
\lim_{n\to\infty}p_1^{(n)}=1.
\]
If $0<m_{12}<m_{21}\leq1$, then $\lim_{n\to\infty}p_1^{(n)}=0$.
\end{theorem}

On large islands the last wild type of the colonization process is therefore on the lower temperature island.

\begin{proof}
Let $d\coloneqq m_{12}-m_{21}$.
Choose integers $q\geq1$ and $K\geq1$ with $qd\geq1$ and $Km_{12}\geq2$.
Let $n\geq4K/d$ and $W\coloneqq\lfloor dn/2\rfloor$.
Let $Q(u,v)$ be the probability of a fixation location on island $2$ from $u$ wild types on island $1$ and $v$ wild types on island $2$.
Then $Q(0,v)=1$ for $v\geq1$ and $Q(u,0)=0$ for $u\geq1$.
The steps without a change of state drop out of the first-step equations.
For $u\geq1$ and $v\geq1$ the first-step equations therefore read
\begin{equation}\label{eq:col-step}
    (\alpha u+\beta v)\,Q(u,v)=\alpha u\,Q(u-1,v)+\beta v\,Q(u,v-1).
\end{equation}

First, we bound the coefficients $\alpha$ and $\beta$ while island $1$ has at most $W$ wild types.
For $u\leq W$ and $v\leq W$, the coefficients satisfy $\alpha\leq(1-d)n$ and $\beta\geq(1+d)(n-W)\geq n$.
Hence $q\alpha\leq(q-1)\beta$.
For $u\leq W$ and every $v$, the coefficients satisfy $\alpha\leq n$ and $\beta\geq m_{12}(n-W)\geq m_{12}n/2$.
Hence $\alpha\leq K\beta$.

Second, we define a potential.
Let $\phi_j\coloneqq qj/(qj-q+1)$ for $j\leq W$ and $\phi_j\coloneqq j/(j-K)$ for $j>W$.
Let $h(v)\coloneqq\phi_1\phi_2\cdots\phi_v$.
Every factor $\phi_j$ is at least $1$, because $W\geq K$.
We show $(u+1)\,Q(u,v)\leq h(v)$ for $u\leq W$ by induction on $u+v$.
At $u=0$ the bound is $1\leq h(v)$.
At $v=0$ the bound is $0\leq h(0)$.
For $u\geq1$ and $v\geq1$, Eq.~\eqref{eq:col-step} and the induction hypothesis reduce the bound to $\alpha\phi_v\leq\beta v(\phi_v-1)$.
For $v\leq W$ the inequality $\alpha\phi_v\leq\beta v(\phi_v-1)$ is equivalent to $q\alpha\leq(q-1)\beta$.
For $v>W$ the inequality is equivalent to $\alpha\leq K\beta$.

Third, every state with $u\geq W$ satisfies $(W+1)\,Q(u,v)\leq h(n)$.
The proof is again an induction on $u+v$, because Eq.~\eqref{eq:col-step} writes $Q(u,v)$ as a weighted average of $Q(u-1,v)$ and $Q(u,v-1)$.
States with $v=0$ have $Q(u,v)=0$.
The states with $u=W$ satisfy the bound by the second step, since $h$ is increasing.

Fourth, we bound $h(n)$.
Bernoulli's inequality gives $(qv+1)^q\geq q^qv^{q-1}(v+1)$.
Induction on $v$ then gives $h(W)^q\leq q^qW^{q-1}$.
For $j>W$, the bound $j-K\geq W+1-K\geq dn/4$ gives $\phi_j=1+K/(j-K)\leq e^{4K/(dn)}$.
Hence $h(n)\leq h(W)\,e^{4K/d}$.

A single initial mutant starts in a state with $u\geq W$.
The third and the fourth steps give
\[
(W+1)^q\,Q^q\leq h(W)^q e^{4Kq/d}\leq q^q(W+1)^{q-1}e^{4Kq/d}
\]
for $Q=1-p_1^{(n)}$.
Hence $(W+1)\,Q^q\leq q^qe^{4Kq/d}$.
Since $W+1>dn/2$, the probability $1-p_1^{(n)}$ tends to $0$ as $n\to\infty$.
Exchanging the two islands gives the case $m_{12}<m_{21}$.
\end{proof}

\subsection{Neutral evolution}
At neutrality, the fixation probability for finite $n$ can be obtained from a martingale.
Let $x_1$ and $x_2$ denote the mutant frequencies on islands $1$ and $2$.
In each time step a parent is chosen uniformly among the $2n$ individuals.
The offspring of an island $1$ parent replaces a uniformly chosen individual of island $2$ with probability $m_{12}$ and of island $1$ otherwise.
The offspring of an island $2$ parent replaces a uniformly chosen individual of island $1$ with probability $m_{21}$ and of island $2$ otherwise.

\begin{lemma}[Neutral drift, Lean \Cref{lean:islands}]\label{lem:neutral-drift}
At $r=1$, for all mutant frequencies $x_1$ and $x_2$,
\[
\mathbb{E}[\Delta x_1\mid x_1,x_2]=\frac{m_{21}}{2n}(x_2-x_1),
\qquad
\mathbb{E}[\Delta x_2\mid x_1,x_2]=\frac{m_{12}}{2n}(x_1-x_2).
\]
Hence $\mathbb{E}[\Delta(m_{12}x_1+m_{21}x_2)\mid x_1,x_2]=0$.
\end{lemma}

\begin{proof}
For $x_1$ there are two kinds of replacement events.
The first kind is replacement within island $1$.
The second kind is replacement of an island $1$ individual by offspring from island $2$.
Thus the expected change $\Delta x_1$ in the island $1$ mutant frequency is
\begin{equation}
    \mathbb{E}[\Delta x_1\mid x_1,x_2]
    =
    \frac{1}{2n}
    \Big[
        (1-m_{12})[x_1(1-x_1)-(1-x_1)x_1]
        +
        m_{21}[x_2(1-x_1)-(1-x_2)x_1]
    \Big].
\end{equation}
The within island contribution cancels under neutrality, leaving
\begin{equation}
    \mathbb{E}[\Delta x_1\mid x_1,x_2]
    =
    \frac{m_{21}}{2n}(x_2-x_1).
\end{equation}
Similarly,
\begin{equation}
    \mathbb{E}[\Delta x_2\mid x_1,x_2]
    =
    \frac{m_{12}}{2n}(x_1-x_2).
\end{equation}
Therefore
\begin{equation*}
    \mathbb{E}[\Delta(m_{12}x_1+m_{21}x_2)\mid x_1,x_2]=0.
\end{equation*}
\end{proof}

\begin{theorem}[Neutral fixation probability on two islands, Lean \Cref{lean:islands}]\label{thm:neutral-islands}
Let $0<m_{12}<1$ and $0<m_{21}<1$.
At $r=1$, the fixation probability from mutant frequencies $x_1$ and $x_2$ is
\[
\rho(x_1,x_2)=\frac{m_{12}x_1+m_{21}x_2}{m_{12}+m_{21}}.
\]
\end{theorem}

\begin{proof}
By \Cref{lem:neutral-drift}, $m_{12}x_1+m_{21}x_2$ is a martingale.
After normalization of the martingale to $0$ at the all wild type state and $1$ at the all mutant state, the function $\rho$ satisfies the first-step equations of the chain.
From every other state the chain reaches the all mutant state with positive probability, because $m_{12}$ and $m_{21}$ are positive.
The first-step equations with the two boundary values therefore have a unique solution, by the maximum principle of \Cref{lean:islands}.
\end{proof}

\begin{corollary}[Initial island, Lean \Cref{lean:islands}]\label{cor:initial-island}
Assume the conditions of \Cref{thm:neutral-islands}.
The fixation probabilities of a single mutant on island $1$ and of a single mutant on island $2$ are
\[
\rho(1/n,0)=\frac{m_{12}}{n(m_{12}+m_{21})},
\qquad
\rho(0,1/n)=\frac{m_{21}}{n(m_{12}+m_{21})}.
\]
The ratio of the two fixation probabilities is $m_{21}/m_{12}$.
\end{corollary}

For $m_{12}=10^{-3}$ and $m_{21}=10^{-1}$ the ratio is $100$.
Conditioning on mutant fixation multiplies the probability of each transition by the ratio of $\rho$ after the transition to $\rho$ before the transition.
The neutral fixation location probabilities of Figure~5 of the main text come from the exact Markov chain on the numbers of mutants.
At $r=1$, the exact computation of Figure~5\textbf{c} of the main text gives $p_1>1/2$ approximately when $|m_{12}-1/2|>|m_{21}-1/2|$.
The migration probabilities with $p_1=1/2$ are therefore close to the lines $m_{12}=m_{21}$ and $m_{12}+m_{21}=1$.
Unlike the colonization boundary $m_{12}=m_{21}$ of \Cref{thm:large-islands}, the neutral boundary is a result of the exact computation alone.

For the two islands with $N=20$, $m_{12}=10^{-3}$, and $m_{21}=10^{-1}$, the state is the pair of numbers of mutants on the two islands.
The distribution at time $t$ is the initial distribution times the $t^\text{th}$ power of the transition matrix of the conditioned chain.
An individual drawn uniformly from island $i$ is a mutant with probability $m_i(t)$, the mean mutant fraction on island $i$.
Every individual on island $i$ is a mutant with probability $c_i(t)$.
A mutant drawn from island $i$ raises the probability of fixation to $f_t/m_i(t)$.
An island $i$ with every individual a mutant raises the probability to $f_t/c_i(t)$.
The threshold fitness for island $2$ as the better island to sample and the unbiased fitness $r^*$ are found by linear interpolation.

\section{Computational methods}\label{sec:computation}

We compute fixation probabilities, fixation locations, and fixation times exactly.
The computation solves linear equations for the Markov chain of the Birth-death process on the mutant sets.
A mutant set $X\subseteq V$ is the set of vertices occupied by mutants.
A graph with $N$ vertices has $2^N$ mutant sets.

In one time step an individual is chosen to reproduce with probability proportional to fitness.
A mutant has fitness $r$ and a wild type has fitness $1$.
The total fitness of the population with mutant set $X$ is $W(X)\coloneqq r\lvert X\rvert+N-\lvert X\rvert$.
The offspring replaces the individual at an out-neighbor of the parent, chosen uniformly at random.
Let $f_X(y)$ be the fitness of the individual at vertex $y$ when the mutant set is $X$.
Let $d^+(y)$ be the number of out-neighbors of $y$.
We call $f_X(y)/d^+(y)$ the weight of the edge from $y$ to an out-neighbor $z$ of $y$.
The probability of a time step with the parent at $y$ and the offspring placed on $z$ is the weight of the edge divided by $W(X)$.

A time step is active if the step changes the mutant set (\Cref{sec:factorization}).
A time step along the edge from $y$ to $z$ is active exactly when the individuals at $y$ and $z$ have different types.
Let $a(X)$ be the sum of the weights of the edges from a mutant to a wild type and of the edges from a wild type to a mutant.
A time step from $X$ is therefore active with probability $a(X)/W(X)$.
The number of time steps from $X$ until the next active step is geometric with mean $W(X)/a(X)$.
Let $Q(X,X')$ be the probability of the mutant set $X'$ after the next active step from $X$.
The probability $Q(X,X')$ equals the sum of the weights of the edges producing $X'$, divided by $a(X)$.
No time step changes the mutant sets $\emptyset$ and $V$.
The two mutant sets are therefore absorbing.
A mutant set with $a(X)=0$ is also absorbing.
We call the other mutant sets transient.

Let $Z(X,v)$ be the probability of fixation with fixation location $v$, starting from the mutant set $X$.
Fixation with fixation location $v$ happens through the active step from $V\setminus\{v\}$ to $V$.
Let $B(X,v)\coloneqq p_v$ if $X=V\setminus\{v\}$ and $B(X,v)\coloneqq0$ otherwise, with $p_v$ the probability of \Cref{prop:factorization}.
Conditioning on the first active step from a transient mutant set $X$ yields
\[
Z(X,v)=B(X,v)+\sum_{X'}Q(X,X')\,Z(X',v),
\]
with the sum over the transient mutant sets $X'$.
Hence $Z=(I-Q)^{-1}B$, with $I$ the identity matrix and $Q$ restricted to the transient mutant sets.
The fixation probability is $\rho(X)=\sum_vZ(X,v)$.
The fixation location probability from the initial vertex $u$ is $L_r^G(v\mid u)=Z(\{u\},v)/\rho(\{u\})$.

The escape count $n_u(X)$ of \Cref{sec:factorization} is the expected number of active steps from $X$, starting from a single mutant at $u$.
Each escape from a transient mutant set $X$ follows either the start or an active step from a transient mutant set $X'$ into $X$.
Taking expectations yields
\[
n_u(X)=\mathbbm1[X=\{u\}]+\sum_{X'}n_u(X')\,Q(X',X),
\]
with the sum over the transient mutant sets $X'$.
Hence $n_u=(I-Q)^{-\top}e_{\{u\}}$, with $e_{\{u\}}$ the indicator vector of the mutant set $\{u\}$.
Conditional on fixation, the expected number of active steps from $X$ is $n_u(X)\,\rho(X)/\rho_r^G(u)$ (\Cref{prop:endgame}).
Before each active step from $X$ the process spends on average $W(X)/a(X)$ time steps with mutant set $X$.
The fixation time from the initial vertex $u$, conditional on fixation, is therefore
\[
T_r^G(u)=\frac{1}{\rho_r^G(u)}\sum_Xn_u(X)\,\rho(X)\,\frac{W(X)}{a(X)},
\]
with the sum over the transient mutant sets $X$.
For a uniformly random initial vertex, the expected fixation time is the average of $T_r^G(u)$ over $u$ with weights $\rho_r^G(u)$.

We solve the linear systems directly with a sparse LU factorization of $I-Q$.
The matrix $I-Q$ has one row for each transient mutant set.
A graph with $N$ vertices has up to $2^N-2$ transient mutant sets.
The cost therefore grows exponentially with $N$.
The solver handles graphs with up to $16$ vertices.
At finite $r$ we are not aware of a polynomial time algorithm for general graphs.
Under colonization, \Cref{prop:colonization-directed} determines the possible fixation locations on every directed graph in polynomial time.
At finite $r$, whether a vertex can be the fixation location is determined exactly by reachability between mutant sets in the same Markov chain.
The sufficient conditions of the section on possible locations of fixation are checked in polynomial time.

The probabilities $f_t$ and $m_w(t)$ of \Cref{sec:watching} depend on the distribution of the mutant set at time step $t$, conditional on fixation.
The Markov chain conditioned on fixation multiplies the probability of each time step from $X$ to $X'$ by $\rho(X')/\rho(X)$.
For Figure~1 of the main text and for the graphs with $N=7$ vertices, the initial mutant set is $\{u\}$ with probability $\rho_r^G(u)/\sum_w\rho_r^G(w)$.
The probability $\rho_r^G(u)/\sum_w\rho_r^G(w)$ is the law of the initial vertex conditional on fixation, for a uniformly random initial vertex.
Multiplying the initial distribution $t$ times by the transition matrix of the conditioned chain yields the distribution at time step $t$.
The probability $f_t$ is the probability of the mutant set $V$ at time step $t$.
The probability $m_w(t)$ is the probability of a mutant set containing $w$ at time step $t$.
The expected fixation time is rounded to the nearest time step.

\section{Lean statements}\label{sec:lean}
We checked many of our mathematical claims with the Lean~4 proof assistant, version \texttt{4.36.0-rc2}, and the \texttt{Mathlib} library.
The Lean files are in the folder \texttt{lean} of the code repository.
\begin{itemize}
\item \texttt{FixLoc/Poly.lean} defines the polynomial $P_N$ of \Cref{lem:cycle-pgf} and proves the recurrence, the first values, and positivity of $P_N$.
\item \texttt{FixLoc/Walk.lean} proves the generating function, the fixation probability, and the exact mean of the walk of the block length.
\item \texttt{FixLoc/Variance.lean} proves the exact variance of $H$ in \Cref{thm:cycle-selection}.
\item \texttt{FixLoc/CLTExpansion.lean}, \texttt{FixLoc/CLT.lean}, and \texttt{FixLoc/CLTLaw.lean} prove the central limit theorem of \Cref{thm:cycle-selection}.
\item \texttt{FixLoc/Concentration.lean} proves \Cref{cor:cycle-region}.
\item \texttt{FixLoc/Fourier.lean} proves \Cref{lem:cycle-pgf} for complex $z$ and the Fourier formula for the fixation location probabilities.
\item {\sloppy \texttt{FixLoc/Neutral.lean}, \texttt{FixLoc/NeutralBound.lean}, and \texttt{FixLoc/NeutralConstant.lean} prove \Cref{prop:cycle-neutral} with the constant of the SI proof.\par}
\item \texttt{FixLoc/Factorization.lean} and \texttt{FixLoc/FinalPhase.lean} prove \Cref{prop:factorization,prop:endgame}.
\item \texttt{FixLoc/Colonization.lean} proves \Cref{prop:colonization-directed,prop:colonization} on the possible fixation locations under colonization.
\item \texttt{FixLoc/Fixability.lean} proves \Cref{thm:fixability-exact,cor:fixability-impossible-basic,thm:refillable-sufficient,prop:gateway,lem:single-in} on the possible fixation locations at finite selection.
\item \texttt{FixLoc/Islands.lean} proves \Cref{lem:neutral-drift,thm:neutral-islands,cor:initial-island} on the two islands.
\item \texttt{FixLoc/LargeIslands.lean} proves \Cref{thm:large-islands} on large islands under colonization.
\item \texttt{FixLoc/Audit.lean} prints the axioms of the headline theorems.
\end{itemize}
The headline theorems depend only on the standard axioms \texttt{propext}, \texttt{Classical.choice}, and \texttt{Quot.sound} of Lean.

The probabilities and expectations of the walk enter the Lean proofs as hypotheses, through the first-step equations of the walk.
The true probabilities and expectations satisfy the first-step equations by the first step analysis in the proof of \Cref{lem:cycle-pgf}.
The probability space of the walk is not constructed in Lean.
The reduction from the Birth-death process on the directed cycle to the walk of the block length (\Cref{sec:directed}) is not formalized.
Each proof has three ingredients.
First, every solution of a second-order linear recurrence with value $0$ at position $0$ is a multiple of one canonical solution (\texttt{eq\_mul\_canon}).
Second, the boundary value problem with values at positions $0$ and $N$ has at most one solution (\texttt{bvp\_unique}).
Third, a closed form satisfies the first-step equations, checked by the tactics \texttt{field\_simp} and \texttt{ring}.
Uniqueness then transfers the closed form to every solution of the first-step equations.
In the Lean files $\alpha$ and $\beta$ are the probabilities of a head advance and of a tail retreat, and \texttt{r\textsuperscript{-1}} is $1/r$.

\clearpage
\subsection{Overview of the formal results}\label{sec:overview}
The table provides the Lean status of every formal mathematical statement of the SI.
The following subsections list the exact Lean statements and the names of the Lean files.

\begin{center}
\footnotesize
\renewcommand{\arraystretch}{1.05}
\begin{tabularx}{\textwidth}{@{}>{\raggedright\arraybackslash}p{3.3cm}>{\raggedright\arraybackslash}X>{\raggedright\arraybackslash}X>{\raggedright\arraybackslash}p{3.2cm}@{}}
\toprule
\textbf{Statement} & \textbf{Summary} & \textbf{Proof ingredients} & \textbf{Lean} \\
\midrule
\Cref{prop:factorization}, Factorization of the fixation location, p.~\pageref{prop:factorization} &
The fixation probability times the fixation location probability of $v$ equals the expected number of active steps from the mutant set $V\setminus\{v\}$ times the probability $p_v$ of a step filling $v$. &
Markov chain on mutant sets. From $V\setminus\{v\}$ every active step fills $v$ with the same probability $p_v$. &
Checked (\Cref{lean:factorization}) \\
\Cref{prop:endgame}, Final phase of fixation, p.~\pageref{prop:endgame} &
The expected number of time steps with one wild type left, conditional on fixation, from the fixation location probabilities and the temperatures. &
Geometric holding times between active steps. Escape counts conditional on fixation. \Cref{prop:factorization}. &
Checked when every vertex also has an incoming edge (\Cref{lean:factorization}) \\
\Cref{thm:fixability-exact}, Exact characterization, p.~\pageref{thm:fixability-exact} &
Fixation at $t$ has positive probability exactly when the configuration graph has a path from $\{s\}$ to $V\setminus\{t\}$ and $t$ has an incoming neighbor other than $t$. &
For $r\ge1$ every legal event has positive probability. Sequences of events are paths in the configuration graph. &
Checked (\Cref{lean:finite}) \\
\Cref{cor:fixability-impossible-basic}, Basic impossibility checks, p.~\pageref{cor:fixability-impossible-basic} &
Fixation at $t\neq s$ is impossible if a vertex is unreachable from $s$ or if $t$ has no incoming neighbor other than $t$. &
\Cref{thm:fixability-exact}. &
Checked (\Cref{lean:finite}) \\
\Cref{thm:refillable-sufficient}, Refillability implies fixability, p.~\pageref{thm:refillable-sufficient} &
A refillable path ending at $t$ makes fixation at $t$ possible. &
An explicit sequence of events. Fill every vertex except $t$ and $u_0$, then move the wild type along the path with refills. &
Checked (\Cref{lean:finite}) \\
\Cref{prop:gateway}, Unidirectional bottlenecks, p.~\pageref{prop:gateway} &
A vertex $t$ on every path from $u$ to some other vertex, and mutant at every time after the first mutant at $t$, is never the fixation location. &
Ancestry of the mutant at $w$ along a directed path through $t$. &
Checked (\Cref{lean:finite}) \\
\Cref{lem:single-in}, Single incoming neighbor, p.~\pageref{lem:single-in} &
A vertex $t\neq u$ with one incoming neighbor keeps a mutant at every later time once the neighbor does. &
The vertex receives offspring only from the neighbor. &
Checked (\Cref{lean:finite}) \\
\Cref{prop:colonization-directed}, Possible fixation locations under colonization, p.~\pageref{prop:colonization-directed} &
Under colonization, $t$ is a possible fixation location exactly when $t\neq u$, every other vertex is reachable from $u$ without $t$, and $t$ has an incoming neighbor. &
Mutants are never replaced. Tracing parents back yields paths avoiding $t$. A breadth-first order proves the converse. One search per vertex yields a polynomial time algorithm. &
Checked (\Cref{lean:col}) \\
\Cref{prop:colonization}, Colonization on undirected graphs, p.~\pageref{prop:colonization} &
On a connected undirected graph under colonization, the impossible fixation locations are the initial vertex and the cut vertices. &
Components of $G\setminus\{t\}$. Special case of \Cref{prop:colonization-directed}. &
Checked (\Cref{lean:col}) \\
Fixation probability on the directed cycle, p.~\pageref{par:block} &
For $r>1$, $\rho=(1-r^{-1})/(1-r^{-N})$. &
First-step equations. Second-order recurrence with two boundary values. &
Checked (\Cref{lean:fix}) \\
\Cref{lem:cycle-pgf}, Exact law of the fixation location, p.~\pageref{lem:cycle-pgf} &
The generating function of the number $H$ of head advances, and the fixation location probabilities as a finite Fourier sum. &
Block of mutants as a random walk. First step analysis. Linear recurrence with distinct roots. Binomial theorem. Roots of unity. &
Checked (\Cref{lean:gf,lean:fourier}) \\
\Cref{prop:cycle-neutral}, Neutral evolution, p.~\pageref{prop:cycle-neutral} &
Under neutral evolution every vertex is the fixation location with probability $1/N+O(N^{-1/2}e^{-\sqrt{N/2}})$. &
\Cref{lem:cycle-pgf} at $r=1$. A lower bound on $|P_N(1-\omega^{\ell})|$ from the binomial theorem, $\sin t\ge2t/\pi$, and $\log(1+s)\ge s/2$. &
Checked (\Cref{lean:neutral}) \\
\Cref{thm:cycle-selection}, Directed cycle under constant selection, p.~\pageref{thm:cycle-selection} &
For fixed $r>1$, the exact mean and the variance of $H$, and a central limit theorem. &
Cumulant generating function from \Cref{lem:cycle-pgf}. Derivatives at $0$. Taylor expansion. Continuity theorem for characteristic functions. &
Checked (\Cref{lean:mean,lean:var,lean:clt}) \\
\Cref{cor:cycle-region}, Concentration of the fixation location, p.~\pageref{cor:cycle-region} &
The fixation location is within order $\sqrt{N}$ vertices of vertex $1+m_N$, near vertex $N/2$ for $r=3$. &
\Cref{thm:cycle-selection}. &
Checked (\Cref{lean:clt}) \\
\Cref{thm:large-islands}, Large islands under colonization, p.~\pageref{thm:large-islands} &
Under colonization $p_1^{(n)}\to1$ as $n\to\infty$ when $m_{12}>m_{21}$, and $p_1^{(n)}\to0$ when $m_{12}<m_{21}$. &
Potential $h(v)/(u+1)$ near fixation. Bernoulli's inequality. &
Checked (\Cref{lean:large-islands}) \\
\Cref{lem:neutral-drift,thm:neutral-islands,cor:initial-island}, Neutral fixation probability on two islands, p.~\pageref{thm:neutral-islands} &
At $r=1$ the fixation probability is $(m_{12}x_1+m_{21}x_2)/(m_{12}+m_{21})$. A single mutant on island $2$ fixes $m_{21}/m_{12}$ times as often as on island $1$. &
Martingale $m_{12}x_1+m_{21}x_2$. Maximum principle. &
Checked (\Cref{lean:islands}) \\
\bottomrule
\end{tabularx}
\end{center}
\clearpage

\subsection{The polynomial $P_N$}\label{lean:poly}
File \texttt{FixLoc/Poly.lean}.
The definition is the polynomial $P_N(y)=\sum_{m}\binom{N}{2m+1}y^{m}$ of \Cref{lem:cycle-pgf}.
The recurrence and the positivity of $P_N$ enter the proofs of the theorems on the directed cycle.
\begin{Verbatim}[frame=single,fontsize=\small,framesep=2mm]
noncomputable def PN (N : ℕ) (y : ℝ) : ℝ :=
  ∑ m ∈ range (N + 1), (N.choose (2 * m + 1) : ℝ) * y ^ m

lemma PN_succ_succ (n : ℕ) (y : ℝ) :
    PN (n + 2) y = 2 * PN (n + 1) y - (1 - y) * PN n y

lemma PN_pos (N : ℕ) (hN : 1 ≤ N) (y : ℝ) (hy : 0 ≤ y) : 0 < PN N y
\end{Verbatim}

\subsection{Exact law of the number of head advances}\label{lean:gf}
File \texttt{FixLoc/Walk.lean}, theorem \texttt{gf\_head\_advances}.
Let $F(x)$ be the expectation of $z^{H}$ on the event of fixation, for the walk started at position $x$.
The structure \texttt{GFSteps} states the first-step equations of $F$.
A head advance multiplies by $z$.
The theorem states $F(1)=(2\alpha z)^{N-1}/P_N(1-4\alpha\beta z)$ for real $z>0$.
Since $F(1)=\rho\,\mathbb{E}[z^{H}]$, the theorem is Eq.~\eqref{eq:cycle-pgf} of \Cref{lem:cycle-pgf} for real $z>0$.
\Cref{lean:fourier} extends the theorem to complex $z$.
\begin{Verbatim}[frame=single,fontsize=\small,framesep=2mm]
structure GFSteps (α β z : ℝ) (N : ℕ) (F : ℕ → ℝ) : Prop where
  zero : F 0 = 0
  top : F N = 1
  step : ∀ k, k + 2 ≤ N → F (k + 1) = α * z * F (k + 2) + β * F k

theorem gf_head_advances (α β z : ℝ) (hα : 0 < α) (hz : 0 < z) (N : ℕ) (hN : 1 ≤ N)
    (F : ℕ → ℝ) (hF : GFSteps α β z N F) :
    F 1 = (2 * α * z) ^ (N - 1) / PN N (1 - 4 * α * β * z)
\end{Verbatim}

\subsection{Complex $z$ and the Fourier formula}\label{lean:fourier}
\begin{sloppypar}
File \texttt{FixLoc/Fourier.lean}, theorems \texttt{gf\_\allowbreak{}head\_\allowbreak{}advances\_\allowbreak{}complex} and \texttt{fourier\_\allowbreak{}location}.
The definitions \texttt{PNc}, \texttt{canonC}, and \texttt{GFStepsC} are the versions of \texttt{PN}, \texttt{canon}, and \texttt{GFSteps} over the complex numbers.
Theorem \texttt{gf\_\allowbreak{}head\_\allowbreak{}advances\_\allowbreak{}complex} states $F(1)=(2\alpha z)^{N-1}/P_N(1-4\alpha\beta z)$ for every complex $z$ and every solution $F$ of the first-step equations.
For $|z|\le1$ the expectation $F(1)=\rho\,\mathbb{E}[z^{H}]$ converges and satisfies the first-step equations.
The theorem is therefore Eq.~\eqref{eq:cycle-pgf} of \Cref{lem:cycle-pgf}.
Theorem \texttt{fourier\_\allowbreak{}location} is Eq.~\eqref{eq:cycle-fourier}.
The function \texttt{q} is the distribution of $H$ conditional on fixation, with $q(h)$ the probability of $H=h$.
The left side is the probability of $H\equiv k-1$ modulo $N$, the probability of fixation location $k$.
The proof uses the orthogonality of the powers of $\omega\coloneqq e^{2\pi i/N}$.
The sum $\sum_{\ell=0}^{N-1}\omega^{\ell j}$ equals $N$ if $N$ divides $j$ and $0$ otherwise.
The proof also exchanges the finite sum over $\ell$ with the infinite sum over $h$.
\end{sloppypar}
\begin{Verbatim}[frame=single,fontsize=\small,framesep=2mm]
structure GFStepsC (α β z : ℂ) (N : ℕ) (F : ℕ → ℂ) : Prop where
  zero : F 0 = 0
  top : F N = 1
  step : ∀ k, k + 2 ≤ N → F (k + 1) = α * z * F (k + 2) + β * F k

theorem gf_head_advances_complex (α β z : ℂ) (hα : α ≠ 0) (N : ℕ) (hN : 1 ≤ N)
    (F : ℕ → ℂ) (hF : GFStepsC α β z N F) :
    F 1 = (2 * α * z) ^ (N - 1) / PNc N (1 - 4 * α * β * z)

theorem fourier_location (N : ℕ) (hN : 1 ≤ N) (q : ℕ → ℝ) (hq : ∀ h, 0 ≤ q h)
    (hs : Summable q) (k : ℕ) (hk1 : 1 ≤ k) (hkN : k ≤ N) :
    ((∑' h, (if h % N = k - 1 then q h else 0) : ℝ) : ℂ) =
      (1 / N) * ∑ ℓ ∈ range N, exp (-(2 * π * I * ℓ * (k - 1) / N)) *
        ∑' h, (q h : ℂ) * exp (2 * π * I * ℓ * h / N)
\end{Verbatim}

\subsection{Fixation probability}\label{lean:fix}
File \texttt{FixLoc/Walk.lean}, theorem \texttt{fixation\_probability}.
Let $h(x)$ be the probability of fixation for the walk started at position $x$.
The structure \texttt{FixSteps} states the first-step equations of $h$.
The theorem states $h(1)=(1-r^{-1})/(1-r^{-N})$ for $r>1$, the fixation probability $\rho$ of \Cref{sec:directed}.
\begin{Verbatim}[frame=single,fontsize=\small,framesep=2mm]
def alpha (r : ℝ) : ℝ := r / (1 + r)

def beta (r : ℝ) : ℝ := 1 / (1 + r)

structure FixSteps (r : ℝ) (N : ℕ) (h : ℕ → ℝ) : Prop where
  zero : h 0 = 0
  top : h N = 1
  step : ∀ k, k + 2 ≤ N → h (k + 1) = alpha r * h (k + 2) + beta r * h k

theorem fixation_probability (r : ℝ) (hr : 1 < r) (N : ℕ) (hN : 1 ≤ N) (h : ℕ → ℝ)
    (hh : FixSteps r N h) : h 1 = (1 - r⁻¹) / (1 - r⁻¹ ^ N)
\end{Verbatim}

\subsection{Exact mean of the number of head advances}\label{lean:mean}
File \texttt{FixLoc/Walk.lean}, theorem \texttt{mean\_head\_advances}.
Let $U(x)$ be the expectation of $H$ on the event of fixation, for the walk started at position $x$.
The structure \texttt{UpSteps} states the first-step equations of $U$.
A head advance from position $k+1$ adds $1$ on the event of fixation, with probability $h(k+2)$ from the new position.
The theorem establishes $\mathbb{E}[H]=U(1)/h(1)$ in the form of Eq.~\eqref{eq:cycle-mean} of \Cref{thm:cycle-selection}, for $r>1$.
\begin{Verbatim}[frame=single,fontsize=\small,framesep=2mm]
structure UpSteps (r : ℝ) (N : ℕ) (h U : ℕ → ℝ) : Prop where
  zero : U 0 = 0
  top : U N = 0
  step : ∀ k, k + 2 ≤ N → U (k + 1) = alpha r * (U (k + 2) + h (k + 2)) + beta r * U k

theorem mean_head_advances (r : ℝ) (hr : 1 < r) (N : ℕ) (hN : 1 ≤ N) (h U : ℕ → ℝ)
    (hh : FixSteps r N h) (hU : UpSteps r N h U) :
    U 1 / h 1 = ((N : ℝ) - 1) * r / (r - 1) - (r + 1) / (r - 1) ^ 2
      + (N : ℝ) * (r + 1) / ((r - 1) * (r ^ N - 1))
\end{Verbatim}

\subsection{Exact variance of the number of head advances}\label{lean:var}
File \texttt{FixLoc/Variance.lean}, theorems \texttt{variance\_head\_advances} and \texttt{varRem\_bounds}.
\begin{sloppypar}
Let $W(x)$ be the expectation of $H^{2}$ on the event of fixation, for the walk started at position $x$.
The structure \texttt{SqSteps} states the first-step equations of $W$.
A head advance turns $H^{2}$ into $(H+1)^{2}=H^{2}+2H+1$, so the equations involve $U$ and $h$.
The closed form \texttt{WClosed} comes from solving the recurrence with the computer algebra system sympy.
The Lean proof checks the closed form against the first-step equations by the tactics \texttt{field\_simp} and \texttt{ring}.
Uniqueness of the boundary value problem transfers the closed form to every solution.
Theorem \texttt{variance\_head\_advances} establishes $\operatorname{Var}[H]=W(1)/h(1)-\bigl(U(1)/h(1)\bigr)^{2}$ exactly, as the leading term of Eq.~\eqref{eq:cycle-var} plus the remainder \texttt{varRem}.
Theorem \texttt{varRem\_bounds} establishes $0\le\texttt{varRem}\le K_r N^{2}r^{-N}$ for every $N\ge1$, with $K_r=r^{2}(r+1)(r^{2}+2r-1)/(r-1)^{5}$.
The two theorems are Eq.~\eqref{eq:cycle-var} of \Cref{thm:cycle-selection}.
\end{sloppypar}
\begin{Verbatim}[frame=single,fontsize=\small,framesep=2mm]
structure SqSteps (r : ℝ) (N : ℕ) (h U W : ℕ → ℝ) : Prop where
  zero : W 0 = 0
  top : W N = 0
  step : ∀ k, k + 2 ≤ N →
    W (k + 1) = alpha r * (W (k + 2) + 2 * U (k + 2) + h (k + 2)) + beta r * W k

def varRem (r : ℝ) (N : ℕ) : ℝ :=
  (N : ℝ) * r⁻¹ ^ N * (r + 1) * ((N : ℝ) * (r ^ 2 - 1) + 2 * r * (1 - r⁻¹ ^ N))
    / ((1 - r⁻¹ ^ N) ^ 2 * (r - 1) ^ 3)

theorem variance_head_advances (r : ℝ) (hr : 1 < r) (N : ℕ) (hN : 1 ≤ N)
    (h U W : ℕ → ℝ) (hh : FixSteps r N h) (hU : UpSteps r N h U)
    (hW : SqSteps r N h U W) :
    W 1 / h 1 - (U 1 / h 1) ^ 2
      = r * (r + 1) / (r - 1) ^ 3 * ((N : ℝ) - 2 * (r + 1) / (r - 1)) + varRem r N

theorem varRem_bounds (r : ℝ) (hr : 1 < r) (N : ℕ) (hN : 1 ≤ N) :
    0 ≤ varRem r N ∧
      varRem r N ≤ r ^ 2 * (r + 1) * (r ^ 2 + 2 * r - 1) / (r - 1) ^ 5
        * ((N : ℝ) ^ 2 * r⁻¹ ^ N)
\end{Verbatim}

\subsection{Central limit theorem and concentration of the fixation location}\label{lean:clt}
\begin{sloppypar}
Files \texttt{FixLoc/\allowbreak{}CLTExpansion.lean}, \texttt{FixLoc/\allowbreak{}CLT.lean}, \texttt{FixLoc/\allowbreak{}CLTLaw.lean}, and \texttt{FixLoc/\allowbreak{}Concentration.lean}.
The function \texttt{p N} is the distribution of $H$ conditional on fixation on the directed cycle of $N$ vertices.
Let $F(x)$ be the expectation of $e^{i\theta H}$ on the event of fixation, for the walk started at position $x$.
The hypothesis \texttt{hF} of theorem \texttt{clt\_\allowbreak{}head\_\allowbreak{}advances\_\allowbreak{}of\_\allowbreak{}steps} states the first-step equations of $F$ for every real $\theta$.
The hypothesis \texttt{hp} states $\mathbb{E}[e^{i\theta H}]=F(1)/\rho$.
Theorem \texttt{gf\_\allowbreak{}head\_\allowbreak{}advances\_\allowbreak{}complex} of \Cref{lean:fourier} derives Eq.~\eqref{eq:cycle-pgf} at $z=e^{i\theta}$ from the two hypotheses.
The definitions \texttt{meanH} and \texttt{varH} are the exact mean of \Cref{lean:mean} and the exact variance of \Cref{lean:var}.
The law \texttt{stdLaw r N (p N)} is the distribution of $(H-\mathbb{E}[H])/\sqrt{\operatorname{Var}[H]}$.
The law \texttt{stdNormal} is the standard normal distribution.
The theorem states convergence in distribution of \texttt{stdLaw r N (p N)} to \texttt{stdNormal} as $N\to\infty$, for every $r>1$.
The theorem is the central limit theorem of \Cref{thm:cycle-selection}.
\end{sloppypar}

The Lean proof differs from the proof of \Cref{thm:cycle-selection} in one point.
The Lean proof expands one root of a quadratic equation in place of the cumulant generating function.
The Lean proof has five steps.
\begin{enumerate}
\item Let $s(\theta)\coloneqq\sqrt{1-4\alpha\beta e^{i\theta}}$ be the principal square root.
Let $B\coloneqq(1-s)/(2\beta)$, $A\coloneqq s/(\alpha e^{i\theta})$, and $R\coloneqq(1-s)/(1+s)$.
The binomial theorem yields $\mathbb{E}[e^{i\theta H}]=A(\theta)B(\theta)^{N}/\bigl(\rho\,(1-R(\theta)^{N})\bigr)$ (\texttt{gf\_factor}).
\item The root $B$ solves $\beta B^{2}-B+\alpha e^{i\theta}=0$ with $B(0)=1$.
Let $w\coloneqq i\theta$ and $\kappa\coloneqq-r(r^{2}+1)/\bigl(2(r-1)^{3}\bigr)$.
Near $\theta=0$ the difference between $B(\theta)$ and $1+rw/(r-1)-\kappa w^{2}$ is at most a constant times $|\theta|^{3}$ (\texttt{Bf\_expansion}).
\item The Taylor bound for the complex logarithm yields $\bigl|\log B(\theta)-rw/(r-1)-\sigma_r^{2}w^{2}/2\bigr|\le C|\theta|^{3}$ near $\theta=0$, with $\sigma_r^{2}=r(r+1)/(r-1)^{3}$ (\texttt{logBf\_expansion}).
\item Let $\sigma_N^{2}\coloneqq\operatorname{Var}[H]$ and $\theta\coloneqq\tau/\sigma_N$ for a real number $\tau$.
The factors $A(\theta)/A(0)$ and $(1-R(0)^{N})/(1-R(\theta)^{N})$ tend to $1$ as $N\to\infty$.
The exponent $N\log B(\theta)-i\tau\,\mathbb{E}[H]/\sigma_N$ tends to $-\tau^{2}/2$, by the exact mean, the exact variance, and the limit $N/\sigma_N^{2}\to1/\sigma_r^{2}$.
The characteristic function of $(H-\mathbb{E}[H])/\sigma_N$ therefore tends to $e^{-\tau^{2}/2}$ (\texttt{clt\_charFun}).
\item L\'evy's continuity theorem in Mathlib (\texttt{ProbabilityMeasure.\allowbreak{}tendsto\_\allowbreak{}of\_\allowbreak{}tendsto\_\allowbreak{}charFun}) implies convergence in distribution (\texttt{clt\_\allowbreak{}head\_\allowbreak{}advances}).
\end{enumerate}
\begin{Verbatim}[frame=single,fontsize=\small,framesep=2mm]
def mOne (r : ℝ) : ℝ := r / (r - 1)

def vOne (r : ℝ) : ℝ := r * (r + 1) / (r - 1) ^ 3

def cc (r : ℝ) : ℝ := 4 * alpha r * beta r

def sq (r θ : ℝ) : ℂ := (1 - (cc r : ℂ) * exp (θ * I)) ^ ((2 : ℂ)⁻¹)

def Bf (r θ : ℝ) : ℂ := (1 - sq r θ) / (2 * (beta r : ℂ))

def qL (r : ℝ) (w : ℂ) : ℂ := ((mOne r : ℝ) : ℂ) * w + ((vOne r / 2 : ℝ) : ℂ) * w ^ 2

theorem logBf_expansion (r : ℝ) (hr : 1 < r) :
    ∃ δ > 0, ∃ C : ℝ, ∀ θ : ℝ, |θ| ≤ δ →
      ‖Bf r θ - 1‖ ≤ 1 / 2 ∧ ‖log (Bf r θ) - qL r (θ * I)‖ ≤ C * |θ| ^ 3

def rhoH (r : ℝ) (N : ℕ) : ℝ := (1 - r⁻¹) / (1 - r⁻¹ ^ N)

def meanH (r : ℝ) (N : ℕ) : ℝ :=
  ((N : ℝ) - 1) * r / (r - 1) - (r + 1) / (r - 1) ^ 2
    + (N : ℝ) * (r + 1) / ((r - 1) * (r ^ N - 1))

def varH (r : ℝ) (N : ℕ) : ℝ :=
  r * (r + 1) / (r - 1) ^ 3 * ((N : ℝ) - 2 * (r + 1) / (r - 1)) + varRem r N

def stdLaw (r : ℝ) (N : ℕ) (p : PMF ℕ) : ProbabilityMeasure ℝ :=
  ⟨p.toMeasure.map (fun n : ℕ => ((n : ℝ) - meanH r N) / √(varH r N)), inferInstance⟩

def stdNormal : ProbabilityMeasure ℝ := ⟨gaussianReal 0 1, inferInstance⟩

theorem clt_head_advances_of_steps (r : ℝ) (hr : 1 < r) (p : ℕ → PMF ℕ)
    (F : ℕ → ℝ → ℕ → ℂ)
    (hF : ∀ N, 2 ≤ N → ∀ θ : ℝ,
      GFStepsC (alpha r) (beta r) (exp ((θ : ℂ) * I)) N (F N θ))
    (hp : ∀ N, 2 ≤ N → ∀ θ : ℝ,
      ∑' n, (((p N n).toReal : ℝ) : ℂ) * exp ((θ : ℂ) * n * I)
        = F N θ 1 / (rhoH r N : ℂ)) :
    Tendsto (fun N : ℕ => stdLaw r N (p N)) atTop (𝓝 stdNormal)
\end{Verbatim}

\begin{sloppypar}
Theorem \texttt{concentration\_\allowbreak{}location} is \Cref{cor:cycle-region}.
The hypothesis \texttt{hgf} is Eq.~\eqref{eq:cycle-pgf} at $z=e^{i\theta}$, divided by the fixation probability $\rho$.
The definition \texttt{mN} is $m_N$, and \texttt{vOne r} is $\sigma_r^{2}$.
The function \texttt{cdf (gaussianReal 0 1)} is the standard normal distribution function $\Phi$.
The theorem establishes the lower bound $2\Phi(a/\sigma_r)-1$ on the limit inferior of the probability of $|H-m_N|\le a\sqrt N$.
Measures in Mathlib have values in $[0,\infty]$.
The event $|H-m_N|\le a\sqrt N$ places the fixation location $1+H$ within $a\sqrt N$ vertices of vertex $1+m_N$ (modulo $N$), as in the proof of \Cref{cor:cycle-region}.
The proof applies the portmanteau theorem of Mathlib to the open intervals $(-c_k,c_k)$ with $c_k\coloneqq(k+1)a/\bigl((k+2)\sigma_r\bigr)$.
The intervals increase to $(-a/\sigma_r,a/\sigma_r)$ as $k\to\infty$.
The symmetry of the standard normal distribution yields the probability $2\Phi(a/\sigma_r)-1$ for the interval $(-a/\sigma_r,a/\sigma_r)$.
\end{sloppypar}
\begin{Verbatim}[frame=single,fontsize=\small,framesep=2mm]
def mN (r : ℝ) (N : ℕ) : ℝ := ((N : ℝ) - 1) * r / (r - 1) - (r + 1) / (r - 1) ^ 2

theorem concentration_location (r : ℝ) (hr : 1 < r) (a : ℝ) (ha : 0 < a) (p : ℕ → PMF ℕ)
    (hgf : ∀ N, 2 ≤ N → ∀ θ : ℝ,
      ∑' n, (((p N n).toReal : ℝ) : ℂ) * Complex.exp ((θ : ℂ) * n * Complex.I)
        = (2 * (alpha r : ℂ) * Complex.exp ((θ : ℂ) * Complex.I)) ^ (N - 1)
          / ((rhoH r N : ℂ) * PNc N
            (1 - 4 * (alpha r : ℂ) * beta r * Complex.exp ((θ : ℂ) * Complex.I)))) :
    ENNReal.ofReal (2 * cdf (gaussianReal 0 1) (a / √(vOne r)) - 1)
      ≤ liminf (fun N : ℕ =>
          (p N).toMeasure {n : ℕ | |(n : ℝ) - mN r N| ≤ a * √(N : ℝ)}) atTop
\end{Verbatim}

\subsection{Neutral evolution on the directed cycle}\label{lean:neutral}
\begin{sloppypar}
Files \texttt{FixLoc/\allowbreak{}Neutral.lean}, \texttt{FixLoc/\allowbreak{}NeutralBound.lean}, and \texttt{FixLoc/\allowbreak{}NeutralConstant.lean}, theorems \texttt{neutral\_\allowbreak{}location\_\allowbreak{}sharp} and \texttt{Cneu\_lt}.
The function \texttt{q} is the distribution of $H$ conditional on fixation.
The function \texttt{h} solves the first-step equations of the fixation probability at $r=1$.
The lemma \texttt{fixation\_\allowbreak{}probability\_\allowbreak{}neutral} establishes $h(1)=1/N$.
The hypothesis \texttt{hgf} is Eq.~\eqref{eq:cycle-pgf} at every power of $\omega=e^{2\pi i/N}$, through the first-step equations with $\alpha=\beta=1/2$.
The definition \texttt{Cneu} is the constant $C=\sum_{\ell\ge1}\sqrt{\ell}\,e^{1-\sqrt{\ell}}$ of the proof of \Cref{prop:cycle-neutral}.
Theorem \texttt{neutral\_\allowbreak{}location\_\allowbreak{}sharp} states $|p_k-1/N|\le16\,C\,N^{-1/2}e^{-\sqrt{N/2}}$ for every $N\ge2$ and every vertex $k$.
Theorem \texttt{Cneu\_lt} states $C<10.6$.
The two theorems are \Cref{prop:cycle-neutral} with the constants of the SI proof.
Theorem \texttt{neutral\_\allowbreak{}location\_\allowbreak{}170} establishes the resulting bound $170\,N^{-1/2}e^{-\sqrt{N/2}}$.
The proof follows the proof of \Cref{prop:cycle-neutral}.
The Lean proof uses the explicit square root $x=\sqrt{\sin\vartheta}\,(1-i)\,e^{i\vartheta/2}$ of $1-\omega^{\ell}$, with $\vartheta=\pi\ell/N$ (\texttt{sqrtOneSub}).
The lemma \texttt{PNc\_\allowbreak{}lower} is the lower bound on $|P_N(1-\omega^{\ell})|$.
The Lean proof of $C<10.6$ bounds the first $30$ terms of $C$ one at a time (\texttt{fC\_le\_1} to \texttt{fC\_le\_30}).
Each of the $30$ bounds uses rational bounds on $\sqrt{\ell}$, the bounds $2.71828<e<2.7182819$, and the Taylor polynomial of the exponential function of degree $7$.
Let $G(x)\coloneqq2e^{1-\sqrt x}\,(x+2\sqrt x+2)$.
The derivative of $G$ is $-\sqrt x\,e^{1-\sqrt x}$.
For $\ell\ge2$ the term $\sqrt{\ell}\,e^{1-\sqrt{\ell}}$ is at most $G(\ell-1)-G(\ell)$ (\texttt{fC\_\allowbreak{}le\_\allowbreak{}Gt\_\allowbreak{}sub}).
The sum of the terms with $\ell>30$ is therefore at most $G(30)$ (\texttt{tail\_\allowbreak{}sum\_\allowbreak{}le}).
The two parts imply $C\le9.55576+0.97638<10.6$.
\end{sloppypar}
\begin{Verbatim}[frame=single,fontsize=\small,framesep=2mm]
lemma PNc_lower (N : ℕ) (θ : ℝ) (h0 : 0 < θ) (hπ : θ < π) :
    (1 + √(Real.sin θ)) ^ N - 1
      ≤ 2 * √2 * √(Real.sin θ) * ‖PNc N (1 - exp ((2 * θ : ℝ) * I))‖

def fC (x : ℝ) : ℝ := √x * Real.exp (1 - √x)

def Cneu : ℝ := ∑' ℓ : ℕ, fC ((ℓ : ℝ) + 1)

theorem Cneu_lt : Cneu < 10.6

theorem neutral_location_sharp (N : ℕ) (hN : 2 ≤ N) (q : ℕ → ℝ) (hq : ∀ n, 0 ≤ q n)
    (hs : Summable q) (h : ℕ → ℝ) (hh : FixSteps 1 N h)
    (hgf : ∀ ℓ < N, ∃ F : ℕ → ℂ,
      GFStepsC (1 / 2) (1 / 2) (exp (2 * π * I * ℓ / N)) N F ∧
      ∑' n, (q n : ℂ) * exp (2 * π * I * ℓ * n / N) = F 1 / (h 1 : ℂ))
    (k : ℕ) (hk1 : 1 ≤ k) (hkN : k ≤ N) :
    |(∑' n, if n % N = k - 1 then q n else 0) - 1 / N|
      ≤ 16 * Cneu / √N * Real.exp (-√(N / 2))
\end{Verbatim}

\subsection{Fixation location, escape counts, and the final phase}\label{lean:factorization}
Files \texttt{FixLoc/Factorization.lean} and \texttt{FixLoc/FinalPhase.lean}.
\begin{sloppypar}
The graph is given by the out-neighbor sets \texttt{out v}, without self-loops.
The event $a\to b$ places an offspring of $a$ on the out-neighbor $b$.
The event has weight $f(a)/d^{+}(a)$, with $f(a)=r$ for a mutant and $f(a)=1$ for a wild type.
The total weight of active steps from $X$ is \texttt{actW}, the number $a(X)$ of \Cref{sec:computation}.
The function \texttt{jumpP} is the probability $Q(X,X')$ of \Cref{sec:computation}.
The transient mutant sets (\texttt{transient}) are the mutant sets other than $\emptyset$ and $V$ with a possible active step.
The hypothesis \texttt{hreach} states that from every transient mutant set the Markov chain can reach an absorbing mutant set.
Theorem \texttt{reach\_\allowbreak{}of\_\allowbreak{}strongly\_\allowbreak{}connected} proves \texttt{hreach} for every strongly connected graph.
The function \texttt{f} is the probability of fixation at $v$, and \texttt{g} is the expected number of active steps from $V\setminus\{v\}$.
Both functions enter through their first-step equations.
Theorem \texttt{factorization} states $f=g\,p_v$ on the transient mutant sets, with $p_v=r\theta_v/(r\theta_v+\mathbbm 1[d^{+}(v)\geq1])$.
For $X=\{u\}$ the theorem is \Cref{prop:factorization}, because $f(\{u\})=\rho_r^G(u)\,L_r^G(v\mid u)$ and $g(\{u\})=n_u(V\setminus\{v\})$.
The proof uses a maximum principle.
If the Markov chain can reach an absorbing mutant set from every transient mutant set, the homogeneous first-step equations have only the zero solution (\texttt{homogeneous\_zero}).

The function \texttt{τ v} of \texttt{FixLoc/FinalPhase.lean} is the expected number of time steps with mutant set $V\setminus\{v\}$ on the event of fixation.
The function enters through its first-step equation for single time steps (\texttt{stepP}).
One time step is an active step with probability $a(X)/W(X)$ and leaves the mutant set unchanged otherwise.
Theorem \texttt{final\_phase} is \Cref{prop:endgame}.
The Lean statement assumes that every vertex has an incoming edge as well as an outgoing edge, because the formula of \Cref{prop:endgame} divides by $\theta_v$.
\end{sloppypar}
\begin{Verbatim}[frame=single,fontsize=\small,framesep=2mm]
theorem factorization (out : V → Finset V) (hloop : ∀ v, v ∉ out v) (r : ℝ)
    (hr : 0 < r)
    (hreach : ∀ X ∈ transient out r,
      AbsorbReach (transient out r) (jumpP out r) X)
    (v : V) (f g : Finset V → ℝ)
    (hf : ∀ X ∈ transient out r,
      f X = (if X = univ.erase v then jumpP out r X univ else 0)
        + ∑ Y ∈ transient out r, jumpP out r X Y * f Y)
    (hg : ∀ X ∈ transient out r,
      g X = (if X = univ.erase v then 1 else 0)
        + ∑ Y ∈ transient out r, jumpP out r X Y * g Y) :
    ∀ X ∈ transient out r, f X = g X *
      (r * theta out v / (r * theta out v + if (out v).Nonempty then 1 else 0))

theorem final_phase (out : V → Finset V) (hloop : ∀ v, v ∉ out v) (r : ℝ)
    (hr : 0 < r) [Nonempty V]
    (hreach : ∀ X ∈ transient out r,
      AbsorbReach (transient out r) (jumpP out r) X)
    (hout : ∀ v, (out v).Nonempty) (hin : ∀ v, 0 < theta out v)
    (h : Finset V → ℝ) (f g τ : V → Finset V → ℝ) (X : Finset V)
    (hX : X ∈ transient out r)
    (hf : ∀ v, ∀ X ∈ transient out r,
      f v X = (if X = univ.erase v then jumpP out r X univ else 0)
        + ∑ Y ∈ transient out r, jumpP out r X Y * f v Y)
    (hg : ∀ v, ∀ X ∈ transient out r,
      g v X = (if X = univ.erase v then 1 else 0)
        + ∑ Y ∈ transient out r, jumpP out r X Y * g v Y)
    (hτ : ∀ v, ∀ X ∈ transient out r,
      τ v X = (if X = univ.erase v then h (univ.erase v) else 0)
        + ∑ Y ∈ transient out r, stepP out r X Y * τ v Y) :
    ∑ v, τ v X / h X = (r * ((Fintype.card V : ℝ) - 1) + 1) / r
      * ∑ v, f v X / h X * h (univ.erase v) / theta out v
\end{Verbatim}

\subsection{Possible fixation locations under colonization}\label{lean:col}
File \texttt{FixLoc/Colonization.lean}, theorems \texttt{possible\_fix\_iff} and \texttt{possible\_fix\_iff\_not\_cut}.
The vertex set \texttt{V} is finite.
The relation \texttt{adj x y} holds when the graph has an edge from $x$ to $y$.
Under colonization only mutants reproduce.
A mutant at $x$ places an offspring on an out-neighbor $y$.
The placement changes the mutant set only if a wild type occupies $y$.
Every such placement has positive probability.
The Lean file models the colonization process by the placements of positive probability.
The Lean file does not construct the probability space of the process.
The relation \texttt{ColReach adj u X} holds for every mutant set $X$ reachable by placements from a single mutant at $u$.
The vertex $t$ is a possible fixation location (\texttt{PossibleFix adj u t}) when the mutant set $V\setminus\{t\}$ is reachable and $t$ has an incoming neighbor other than $t$.
The relation \texttt{avoid adj t} is the edge relation of the graph $G\setminus\{t\}$.
The relation \texttt{ReflTransGen R} is the reflexive transitive closure of a relation \texttt{R}.
The condition \texttt{ReflTransGen (avoid adj t) u v} states $v\in R_{G\setminus\{t\}}(u)$.
A vertex $t$ is a cut vertex (\texttt{IsCutVertex adj t}) when two vertices other than $t$ have no path between each other in $G\setminus\{t\}$.

Theorem \texttt{possible\_fix\_iff} is \Cref{prop:colonization-directed}.
Theorem \texttt{possible\_fix\_iff\_not\_cut} is \Cref{prop:colonization}.
The hypothesis \texttt{hsymm} states an undirected graph through a symmetric edge relation.
The hypothesis \texttt{hconn} states a connected graph.
The Lean statement of \Cref{prop:colonization} has no hypothesis $N\geq2$, because the condition $t\neq u$ already requires two vertices.
The proof of the forward direction traces the parents back, by induction over the placements.
The proof of the converse places, at each step, an offspring along an edge leaving the current mutant set in $G\setminus\{t\}$.
The induction is on the number of vertices of $V\setminus\{t\}$ outside the mutant set.
The proof for undirected graphs reverses paths and uses an edge from $G\setminus\{t\}$ into $t$.
\begin{Verbatim}[frame=single,fontsize=\small,framesep=2mm]
inductive ColReach (adj : V → V → Prop) (u : V) : Finset V → Prop
  | start : ColReach adj u {u}
  | grow {X : Finset V} {x y : V} :
      ColReach adj u X → x ∈ X → adj x y → y ∉ X → ColReach adj u (insert y X)

def PossibleFix (adj : V → V → Prop) (u t : V) : Prop :=
  ColReach adj u (univ.erase t) ∧ ∃ x, x ≠ t ∧ adj x t

def avoid (adj : V → V → Prop) (t : V) (a b : V) : Prop :=
  adj a b ∧ a ≠ t ∧ b ≠ t

def IsCutVertex (adj : V → V → Prop) (t : V) : Prop :=
  ∃ a b, a ≠ t ∧ b ≠ t ∧ ¬ ReflTransGen (avoid adj t) a b

theorem possible_fix_iff (adj : V → V → Prop) (u t : V) :
    PossibleFix adj u t ↔
      t ≠ u ∧ (∀ v, v ≠ t → ReflTransGen (avoid adj t) u v) ∧ ∃ x, x ≠ t ∧ adj x t

theorem possible_fix_iff_not_cut (adj : V → V → Prop)
    (hsymm : ∀ a b, adj a b → adj b a) (hconn : ∀ a b, ReflTransGen adj a b) (u t : V) :
    PossibleFix adj u t ↔ t ≠ u ∧ ¬ IsCutVertex adj t
\end{Verbatim}

\subsection{Possible fixation locations at finite selection}\label{lean:finite}
File \texttt{FixLoc/Fixability.lean}.
At finite fitness every Birth-death event changing the mutant set has positive probability.
The relation \texttt{BDStep adj X Y} is the edge relation of the configuration graph $\mathcal C(G)$ of \Cref{thm:fixability-exact}.
A gain places the offspring of a mutant on a wild type.
A loss places the offspring of a wild type on a mutant.
The relation \texttt{FixableAt adj s t} states a sequence of events from the single mutant at $s$ to a mutant set $X$ without $t$, followed by an event from $X$ to the mutant set $V$.
The Lean file models the process by the events of positive probability.
The Lean file does not construct the probability space of the process.
A trajectory is a sequence of mutant sets $X_0,\ldots,X_m$ indexed by time, with one event between consecutive mutant sets.
The relation \texttt{StaysMutant adj u t} is condition (ii) of \Cref{prop:gateway} for the vertex $t$.

\begin{sloppypar}
Theorems \texttt{fixable\_\allowbreak{}at\_\allowbreak{}iff} and \texttt{bdstep\_\allowbreak{}erase\_\allowbreak{}univ\_\allowbreak{}iff} are \Cref{thm:fixability-exact}.
Theorems \texttt{not\_\allowbreak{}fixable\_\allowbreak{}of\_\allowbreak{}unreachable} and \texttt{not\_\allowbreak{}fixable\_\allowbreak{}of\_\allowbreak{}no\_\allowbreak{}in} are the two checks of \Cref{cor:fixability-impossible-basic}.
Theorem \texttt{refillable\_\allowbreak{}fixable} is \Cref{thm:refillable-sufficient}.
The structure \texttt{Refillable} is \Cref{def:refillable}, with the path as a function \texttt{u} on $\{0,\ldots,k\}$ with distinct values.
The Lean statements of \Cref{cor:fixability-impossible-basic,thm:refillable-sufficient} need no hypothesis $s\neq t$.
The Lean statement of \Cref{thm:refillable-sufficient} also covers paths with $k=0$.
Theorems \texttt{gateway} and \texttt{gateway\_\allowbreak{}initial} are \Cref{prop:gateway}.
Condition (i) of \Cref{prop:gateway} is the hypothesis \texttt{hi}, the absence of a path from $u$ to $w$ in $G\setminus\{t\}$.
Theorem \texttt{single\_\allowbreak{}in} is \Cref{lem:single-in}.
\end{sloppypar}

The proof of \Cref{thm:refillable-sufficient} first colonizes every vertex reachable from $s$ in $G\setminus\{u_0\}$.
If $t$ then holds a wild type, the witness $w$ refills $u_0$ directly.
If $t$ holds a mutant, the wild type moves from $u_0$ along the path with refills, as in the proof of \Cref{thm:refillable-sufficient}.
The proof of \Cref{prop:gateway} uses an ancestry lemma.
As long as $t$ has never held a mutant, every mutant is reachable from $u$ in $G\setminus\{t\}$.
The proof of \Cref{lem:single-in} takes the first time a mutant occupies $t$.
\begin{Verbatim}[frame=single,fontsize=\small,framesep=2mm]
inductive BDStep (adj : V → V → Prop) : Finset V → Finset V → Prop
  | gain {X : Finset V} {a b : V} :
      a ∈ X → b ∉ X → adj a b → BDStep adj X (insert b X)
  | loss {X : Finset V} {a b : V} :
      a ∉ X → b ∈ X → adj a b → BDStep adj X (X.erase b)

def FixableAt (adj : V → V → Prop) (s t : V) : Prop :=
  ∃ X, ReflTransGen (BDStep adj) {s} X ∧ BDStep adj X univ ∧ t ∉ X

theorem fixable_at_iff (adj : V → V → Prop) (s t : V) :
    FixableAt adj s t ↔
      ReflTransGen (BDStep adj) {s} (univ.erase t) ∧ ∃ x, x ≠ t ∧ adj x t

theorem bdstep_erase_univ_iff (adj : V → V → Prop) (t : V) :
    BDStep adj (univ.erase t) univ ↔ ∃ x, x ≠ t ∧ adj x t

theorem not_fixable_of_unreachable (adj : V → V → Prop) (s t v : V)
    (hv : ¬ ReflTransGen adj s v) : ¬ FixableAt adj s t

theorem not_fixable_of_no_in (adj : V → V → Prop) (s t : V)
    (ht : ∀ x, adj x t → x = t) : ¬ FixableAt adj s t

structure Refillable (adj : V → V → Prop) (s t : V) (k : ℕ) (u : ℕ → V)
    (w : V) : Prop where
  path : ∀ i < k, adj (u i) (u (i + 1))
  last : u k = t
  inj : ∀ i j, i ≤ k → j ≤ k → u i = u j → i = j
  start : u 0 ≠ s
  witness : adj w (u 0) ∧ ∀ i ≤ k, w ≠ u i
  reach : ∀ v, v ≠ t → v ≠ u 0 → ReflTransGen (avoid adj (u 0)) s v

theorem refillable_fixable (adj : V → V → Prop) (s t : V) (k : ℕ) (u : ℕ → V)
    (w : V) (hP : Refillable adj s t k u w) : FixableAt adj s t

def StaysMutant (adj : V → V → Prop) (u t : V) : Prop :=
  ∀ (X : ℕ → Finset V) (m : ℕ), X 0 = {u} →
    (∀ i < m, BDStep adj (X i) (X (i + 1))) → X m = univ →
    ∀ i j, i ≤ j → j ≤ m → t ∈ X i → t ∈ X j

theorem gateway (adj : V → V → Prop) (u t w : V) (hwt : w ≠ t)
    (hi : ¬ ReflTransGen (avoid adj t) u w) (hii : StaysMutant adj u t) :
    ¬ FixableAt adj u t

theorem gateway_initial (adj : V → V → Prop) (u : V) (hii : StaysMutant adj u u) :
    ¬ FixableAt adj u u

theorem single_in (adj : V → V → Prop) (u t x : V) (htu : t ≠ u)
    (hin : ∀ y, adj y t ↔ y = x) (hx : StaysMutant adj u x) : StaysMutant adj u t
\end{Verbatim}

\subsection{Neutral fixation probability on two islands}\label{lean:islands}
File \texttt{FixLoc/Islands.lean}.
The state $(a,b)$ counts the mutants on island $1$ and on island $2$, with $0\leq a,b\leq n$.
The functions \texttt{up1} and \texttt{dn1} are the probabilities of a gain and of a loss of one mutant on island $1$ in one time step at $r=1$.
The functions \texttt{up2} and \texttt{dn2} are the same probabilities for island $2$.
The function \texttt{stay} is the probability of a time step without change.
The function \texttt{P} is the transition matrix of the chain.
The transient states (\texttt{T}) are the states other than $(0,0)$ and $(n,n)$.
Lemmas \texttt{drift1} and \texttt{drift2} are the two drift formulas of \Cref{lem:neutral-drift} for the mutant counts.
Lemma \texttt{harmonic} states that \texttt{h}, equal to $(m_{12}a+m_{21}b)/\bigl(n(m_{12}+m_{21})\bigr)$, has zero expected change at every state.
Lemma \texttt{reach} proves that from every transient state the chain reaches $(n,n)$ with positive probability, for $0<m_{12}<1$ and $0<m_{21}<1$.
Theorem \texttt{neutral\_fixation} is \Cref{thm:neutral-islands}.
The fixation probability enters through its first-step equations, as in \Cref{lean:factorization}.
The proof applies the maximum principle \texttt{homogeneous\_zero} to the difference of two solutions of the first-step equations.
Theorem \texttt{neutral\_initial\_island} is \Cref{cor:initial-island}.
\begin{Verbatim}[frame=single,fontsize=\small,framesep=2mm]
theorem neutral_fixation (hn : 0 < n) (h12 : 0 < m₁₂) (h12' : m₁₂ < 1)
    (h21 : 0 < m₂₁) (h21' : m₂₁ < 1) (ρ : ℤ × ℤ → ℝ)
    (hρ : ∀ x ∈ T n, ρ x = P n m₁₂ m₂₁ x ((n : ℤ), (n : ℤ))
      + ∑ y ∈ T n, P n m₁₂ m₂₁ x y * ρ y) :
    ∀ x ∈ T n, ρ x = (m₁₂ * x.1 + m₂₁ * x.2) / (n * (m₁₂ + m₂₁))
\end{Verbatim}

\subsection{Large islands under colonization}\label{lean:large-islands}
File \texttt{FixLoc/LargeIslands.lean}.
The state $(u,v)$ counts the wild types on island $1$ and on island $2$.
The functions \texttt{pA} and \texttt{pB} are the probabilities of a step replacing a wild type on island $1$ and on island $2$ in the colonization process.
The structure \texttt{FirstStep} states the boundary values and the first-step equations of the probability of a fixation location on island $1$.
The probability enters through the first-step equations, as in \Cref{lean:factorization}.
Lemma \texttt{ratio} is Eq.~\eqref{eq:col-step}.
Lemmas \texttt{r1\_bound} and \texttt{r2\_bound} are the first step of the proof of \Cref{thm:large-islands}.
Theorems \texttt{region} and \texttt{entry} are the second and the third steps.
Lemmas \texttt{pot\_pow\_le} and \texttt{pot\_le\_exp} are the fourth step.
Theorem \texttt{colonization\_large\_islands} is \Cref{thm:large-islands} for $m_{12}>m_{21}$.
Theorem \texttt{colonization\_large\_islands\_symm} is the case $m_{12}<m_{21}$.
\begin{Verbatim}[frame=single,fontsize=\small,framesep=2mm]
theorem colonization_large_islands {m₁₂ m₂₁ : ℝ} (h21 : 0 < m₂₁)
    (hlt : m₂₁ < m₁₂) (h12 : m₁₂ ≤ 1) (p : ℕ → ℕ → ℕ → ℝ)
    (hp : ∀ n, FirstStep n m₁₂ m₂₁ (p n)) :
    Tendsto (fun n => p n (n - 1) n) atTop (𝓝 1)
      ∧ Tendsto (fun n => p n n (n - 1)) atTop (𝓝 1)
\end{Verbatim}

\bibliographystyle{unsrt}
\bibliography{sources}


\begin{figure}
  \centering
  \includegraphics[width=1\textwidth]{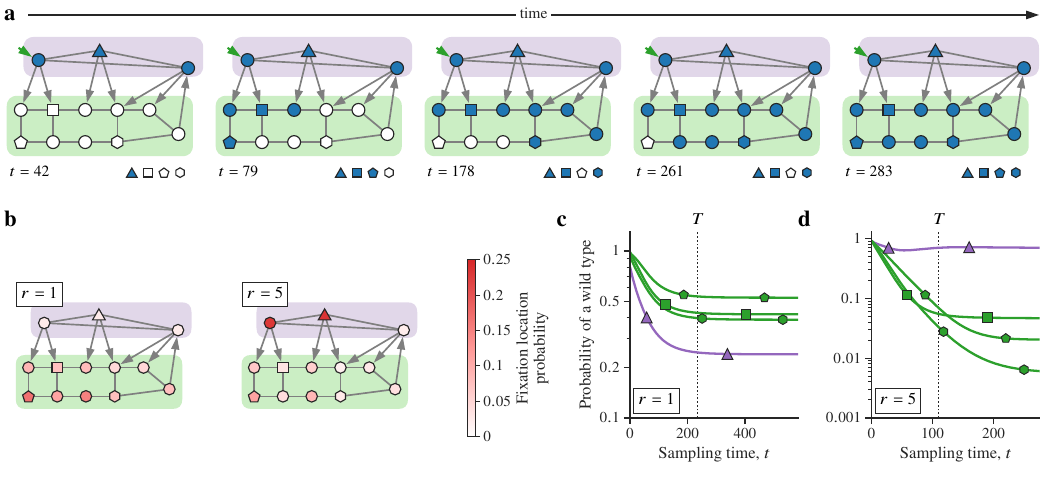}
  \caption{
    The likeliest fixation location is nearly always the best place to sample.
    \textbf{a,} A population structure with $N=13$ vertices.
    The region shaded purple has three vertices.
    Bidirectional edges join each pair of vertices in the purple region.
    The region shaded green has the remaining ten vertices.
    Bidirectional edges join the vertices in the green region.
    The purple region has six directed edges to the green region, and the green region has one directed edge to the purple region.
    The population is displayed at five times over one fixation trajectory at $r=1$, a run of the process ending in fixation.
    The initial mutant arises at a vertex in the purple region (green arrow).
    Vertices occupied by a mutant are blue.
    Vertices occupied by a wild type are white.
    An observer samples the four vertices drawn as a triangle, square, pentagon, and hexagon.
    The four small polygons below each snapshot display the corresponding locations seen by the observer at the four sampled vertices.
    The last wild type occupies the pentagon vertex until fixation at time step $t=283$.
    \textbf{b,} Fixation location probability of each vertex at $r=1$ and $r=5$.
    The initial vertex is uniformly random.
    The probabilities are conditional on fixation.
    The likeliest fixation location is in the green region at $r=1$ and in the purple region at $r=5$.
    \textbf{c,d,} Probability of a wild type given no fixation by the sampling time, against the sampling time, at $r=1$ (\textbf{c}) and $r=5$ (\textbf{d}).
    The vertical axes are logarithmic.
    Each curve is one sampled vertex and has the polygon of the vertex as a marker.
    The purple curve is the sampled vertex in the purple region.
    The green curves are the sampled vertices in the green region.
    The pentagon vertex is the likeliest fixation location at $r=1$.
    The triangle vertex is the likeliest fixation location at $r=5$.
    The dotted line labeled $T$ is the expected fixation time.
  }
  \label{fig:watch}
\end{figure}

\begin{figure}
  \centering
  \includegraphics[width=1\textwidth]{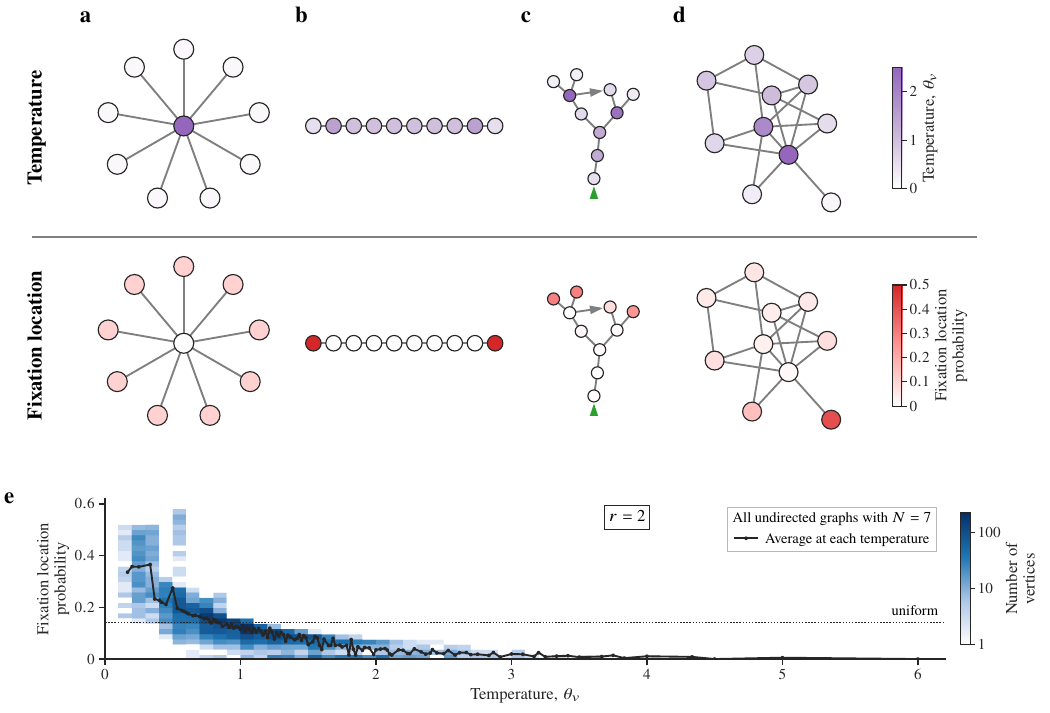}
  \caption{
    The last wild type tends to survive on low temperature vertices.
    \textbf{a} to \textbf{d,} Four population structures with $N=10$ vertices at $r=2$.
    Each structure is drawn once in the top row and once in the bottom row.
    In the top row, vertex shade is the temperature $\theta_v$.
    In the bottom row, vertex shade is the fixation location probability of the vertex.
    \textbf{a,} A star with $N=10$ vertices.
    The temperature of the hub, $9$, exceeds the color scale.
    \textbf{b,} A path with $N=10$ vertices.
    \textbf{c,} A tree with $N=10$ vertices and bidirectional edges, with one added directed edge (arrow).
    The initial mutant is at the vertex marked by the green arrow.
    \textbf{d,} An irregular graph with $N=10$ vertices.
    In \textbf{a}, \textbf{b}, and \textbf{d} the initial vertex is uniformly random.
    The probabilities are conditional on fixation.
    \textbf{e,} Heatmap of the number of vertices versus the fixation location probability and the temperature at $r=2$, for every vertex of every connected undirected graph with $N=7$ vertices ($5{,}971$ vertices of $853$ graphs).
    The initial vertex is uniformly random.
    The black curve connects the average fixation location probability of the vertices at each distinct temperature.
    The dotted line labeled uniform is the value $1/7$ of a uniform fixation location.
  }
  \label{fig:rare}
\end{figure}

\begin{figure}
  \centering
  \includegraphics[width=1\textwidth]{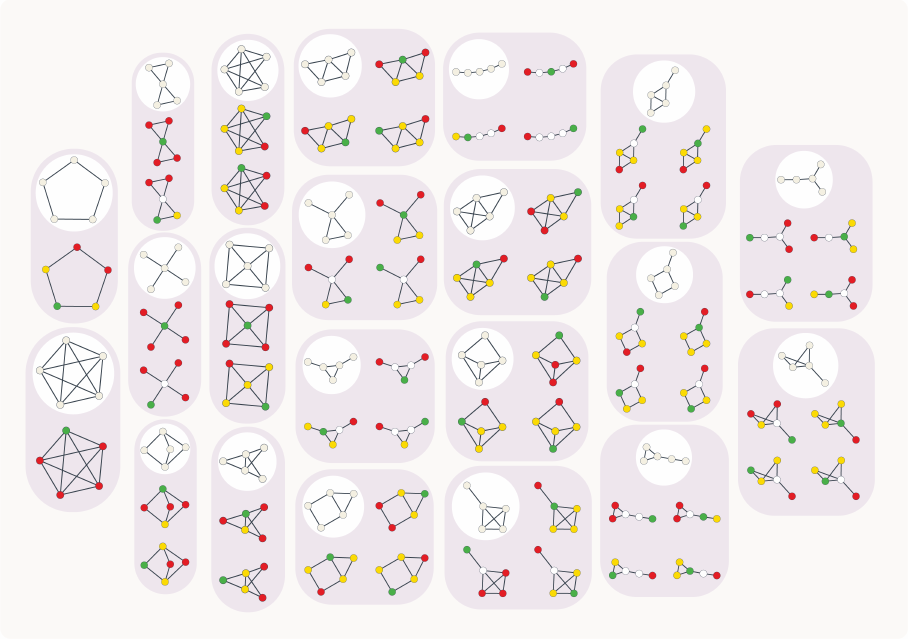}
  \caption{
    Fixation location probabilities for all $21$ connected undirected graphs with $N=5$ vertices, for $r\to\infty$.
    Each shaded rounded box contains one graph.
    The white circle in each box displays the graph with off-white vertices.
    The colored copies of the graph display every non-isomorphic initial mutant location.
    The initial mutant location is green.
    The locations with the highest fixation location probability are red.
    Several locations are red when the locations tie for the highest probability.
    Impossible fixation locations are white.
    The impossible fixation locations other than the initial mutant location are exactly the cut vertices, the vertices whose removal disconnects the graph.
    All other possible fixation locations are yellow.
    We provide a polynomial time algorithm to determine the possible fixation locations of the colonization process on general directed graphs (see \SI{} for details).
  }
  \label{fig:examples-n5} 
\end{figure}

\begin{figure}
  \centering
  \includegraphics[width=0.70\textwidth]{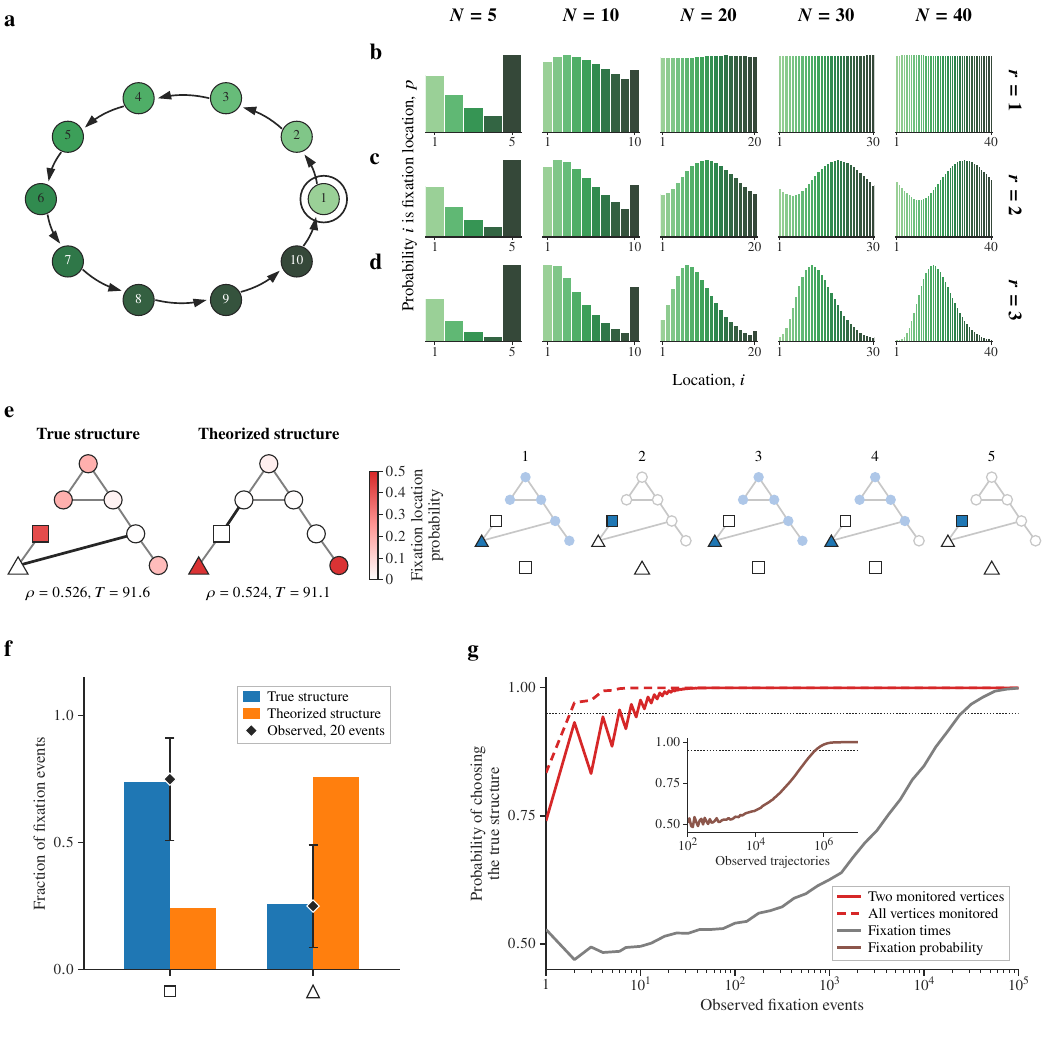}
  \caption{
    The fixation location is not always the farthest vertex from the initial mutant.
    The fixation location distinguishes population structures.
    \textbf{a} to \textbf{d,} Directed cycle populations.
    Location $i$ has a directed edge to location $i+1$.
    Location $N$ has a directed edge to location $1$.
    Initially a single mutant occupies location $i=1$.
    The remaining population consists of wild types.
    \textbf{a,} Directed cycle with $N=10$.
    \textbf{b,c,d,} Numerical computation of the fixation location probability $p$ for $r=1$, $2$, and $3$, in order.
    Population sizes $N=5$, $10$, $20$, $30$, and $40$ are displayed from left to right.
    The scaling of each plot differs.
    The sum of the probabilities in each plot is $1$.
    When $r=1$ and $N$ is large, the fixation location is roughly uniformly distributed.
    When $r=3$ and $N$ is large, the average fixation location approaches $N/2$.
    See \SI{} for details and proofs.
    \textbf{e} to \textbf{g,} Structures with nearly equal fixation probabilities and fixation times are distinguished by fixation locations.
    \textbf{e,} A true structure and a theorized structure on a common set of $N=7$ vertices, at $r=2$.
    The black edge has different positions in the two structures.
    Vertex shade is the fixation location probability of the vertex for a uniformly random initial vertex.
    The fixation probabilities $\rho$ and the fixation times $T$ of the two structures are within $1\%$ of each other.
    An observer monitors only the two vertices drawn as a square and a triangle.
    The five small graphs display five fixation events simulated on the true structure.
    Each small graph displays the population at the last time step with a wild type at a monitored vertex.
    Wild types are white.
    Mutants are blue.
    The observer cannot see the other vertices (drawn as circles).
    The shape below each event is the monitored vertex last occupied by a wild type.
    We call the monitored vertex last occupied by a wild type the observation of the fixation event.
    The observer monitors the square vertex and the triangle vertex long enough for fixation to have occurred with probability at least $0.95$ (see \SI{}).
    \textbf{f,} Fraction of fixation events with the square vertex or the triangle vertex as the monitored vertex last occupied by a wild type.
    Bars are the predictions of the true structure (blue) and of the theorized structure (orange).
    Diamonds are the observed fractions in $20$ fixation events simulated on the true structure, including the five events of \textbf{e}.
    Error bars are $95\%$ confidence intervals for the two fractions.
    \textbf{g,} Probability of choosing the true structure against the number of observed fixation events.
    For the solid red curve, the observer notes the monitored vertex last occupied by a wild type after each fixation event.
    For the dashed red curve, the observer monitors every vertex and notes the fixation location.
    For the gray curve, the observer notes the fixation time.
    For the brown curve in the inset, the observer notes whether each trajectory ends in fixation or extinction.
    Each structure assigns a probability to each possible observation.
    For the solid red curve, the probabilities are the bars of \textbf{f}.
    Under each structure, the probability of all observations so far is the product of the probabilities of the individual observations.
    The observer chooses the structure with the larger probability of all observations so far.
    With two monitored vertices, the probability of choosing the true structure first exceeds $0.95$ after $6$ observed fixation events.
    With every vertex monitored, the probability first exceeds $0.95$ after $2$ observed fixation events.
    For the fixation time, the probability first exceeds $0.95$ after about $25{,}000$ observed fixation events.
    In the inset, the probability first exceeds $0.95$ after about $560{,}000$ observed trajectories.
    The dotted lines are $0.95$.
  }
  \label{fig:test}
\end{figure}

\begin{figure}
  \centering
  \includegraphics[width=0.88\textwidth]{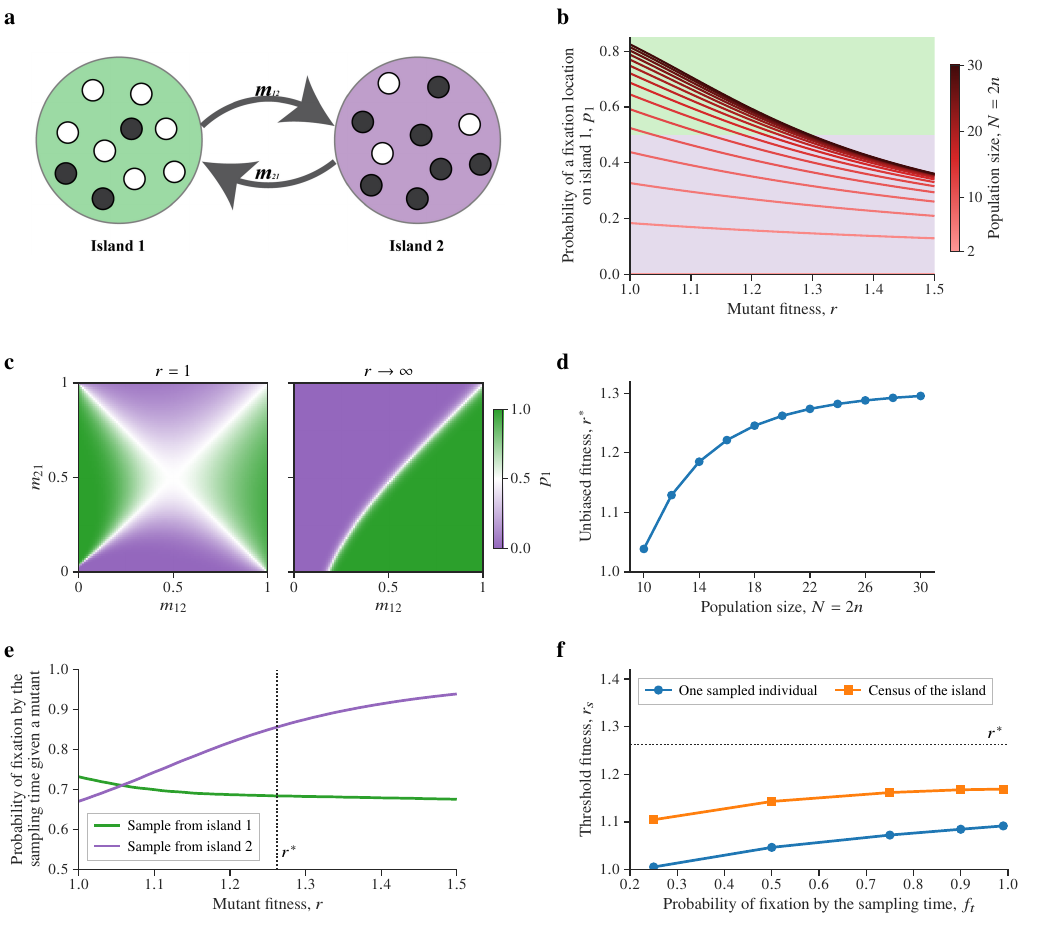}
  \caption{
    The more likely island of the fixation location depends on the mutant fitness, the migration probabilities, and the population size.
    There are two islands, island $1$ (green) and island $2$ (purple).
    Each island has population size $n$.
    The total population size is $N=2n$.
    Each island is a well-mixed population.
    Migration between the islands occurs with probabilities $m_{12}$ and $m_{21}$.
    When an individual on island $1$ reproduces, the offspring is placed on island $2$ with probability $m_{12}$ and remains on island $1$ with probability $1-m_{12}$.
    When an individual on island $2$ reproduces, the offspring is placed on island $1$ with probability $m_{21}$ and remains on island $2$ with probability $1-m_{21}$.
    Conditional on the destination island, the replaced individual is chosen uniformly from the destination island.
    An individual on island $i$ has an edge of weight $(1-m_{ij})/n$ to each individual on island $i$, including itself.
    The individual has an edge of weight $m_{ij}/n$ to each individual on the other island $j$.
    Initially island $1$ has one mutant and $n-1$ wild types.
    Island $2$ initially has $n$ wild types.
    \textbf{a,} Example of a two island population with $n=10$ and $N=20$.
    Island $1$ has $3$ mutants.
    Island $2$ has $7$ mutants.
    Wild types are white.
    Mutants are dark gray.
    \textbf{b,} The probability $p_1$ of a fixation location on island $1$ against the mutant fitness.
    Each curve is a value of $N$.
    Here $m_{12}=10^{-3}$ and $m_{21}=10^{-1}$.
    For $p_1$ above $1/2$ (region shaded green), the fixation location is more likely on island $1$.
    For $p_1$ below $1/2$ (region shaded purple), the fixation location is more likely on island $2$.
    For $p_1=1/2$, the fixation location is unbiased.
    \textbf{c,} For $N=20$, the probability $p_1$ of a fixation location on island $1$ against the migration probabilities $m_{12}$ and $m_{21}$.
    The two panels are $r=1$ and $r\to\infty$, in order.
    \textbf{d,} The unbiased fitness $r^*$ against the population size $N$.
    The unbiased fitness is the mutant fitness with $p_1=1/2$.
    For $N\leq8$, $p_1$ is below $1/2$ for every $r$ from $1$ to $1.5$.
    Here $m_{12}=10^{-3}$ and $m_{21}=10^{-1}$.
    \textbf{e,} An observer samples one individual uniformly from island $1$ (green) or from island $2$ (purple).
    The sampling time is the number of time steps from the initial mutant to the sample.
    In \textbf{e}, the sampling time is the fixation time $T$ at each $r$.
    The curves are the probability of fixation by the sampling time, given a mutant in the sample.
    All probabilities in \textbf{e} and \textbf{f} are conditional on fixation.
    The parameters are $N=20$, $m_{12}=10^{-3}$, and $m_{21}=10^{-1}$.
    The dotted line is the unbiased fitness $r^*=1.26$.
    Below $r=1.06$, the probability of fixation by the sampling time is higher given a mutant sampled from island $1$ than given a mutant sampled from island $2$.
    Above $r=1.06$, the probability of fixation by the sampling time is higher given a mutant sampled from island $2$ than given a mutant sampled from island $1$.
    \textbf{f,} The threshold fitness $r_s$ against the probability $f_t$ of fixation by the sampling time.
    In \textbf{f}, the sampling time is the first time step with $f_t$ at least the value on the horizontal axis.
    Above $r_s$, island $2$ is the better island to sample.
    Circles are one sampled individual.
    Squares are a census of the whole island.
    The dotted line is $r^*$.
  }
  \label{fig:islands}
\end{figure}